\documentclass[preprint,aps,pre,superscriptaddress,citeautoscript,floatfix,longbibliography]{revtex4-1}
\usepackage{graphics,graphicx,mathrsfs,color,amsmath,appendix,framed,tcolorbox,bm,enumerate}
\usepackage[colorlinks]{hyperref}
\usepackage{empheq}
\usepackage{cases}
\newcommand{\ud}{\mathrm{d}}

\begin{document}

\title{Existent condition of partially wet state in capillary tubes}

\author{Chen Zhao}
\affiliation{South China Advanced Institute for Soft Matter Science and Technology, School of Emergent Soft Matter, South China University of Technology, Guangzhou 510640, China}

\author{Jiajia Zhou}
\affiliation{South China Advanced Institute for Soft Matter Science and Technology, School of Emergent Soft Matter, South China University of Technology, Guangzhou 510640, China}
\affiliation{Guangdong Provincial Key Laboratory of Functional and Intelligent Hybrid Materials and Devices, South China University of Technology, Guangzhou 510640, China}
\affiliation{State Key Laboratory of Pulp and Paper Engineering, South China University of Technology, Guangzhou 510640, China}

\author{Masao Doi}
\affiliation{Wenzhou Key Laboratory of Biomaterials and Engineering, Wenzhou Institute, University of Chinese Academy of Sciences, Wenzhou 325000, China}
\affiliation{Oujiang Laboratory (Zhejiang Lab for Regenerative Medicine, Vision and Brain Health), Wenzhou 325000, China}

\date{\today}

\begin{abstract} 
We develop a theory that predicts the equilibrium states of a fluid contained in a capillary which has corners. Each section of the tube can take  three states: completely wet state where the tube section is completely occupied by the fluid, partially wet state where only the corners are occupied by the fluid known as corner film or finger, and completely dry state.  We calculate the phase diagram of these states for a square tube with rounded corners. It is shown that the partially wet state can exist only in a certain region in the parameter space spanned by the equilibrium contact angle and the corner curvature.
\end{abstract}

\maketitle


The phenomena of wetting and drying in porous media are not only ubiquitous in everyday life, but also play a vital role in a wide range of applications \cite{Bear, dBQ, deGennes1985, Bonn2009}.
This spans diverse fields such as environmental science, oil recovery, and various industrial processes of food, textiles, and pharmaceuticals.
A key quantity that characterizes the wetting state is the saturation, $s$, defined as the volume of wetting fluid divided by the volume available to the fluid.
Based on this value, the medium can be divided into regions of distinct wetting states: completely wet ($s=1$), partially wet ($0<s<1$), and dry ($s=0$).

\begin{figure}[ht]
  \includegraphics[width=0.9\columnwidth]{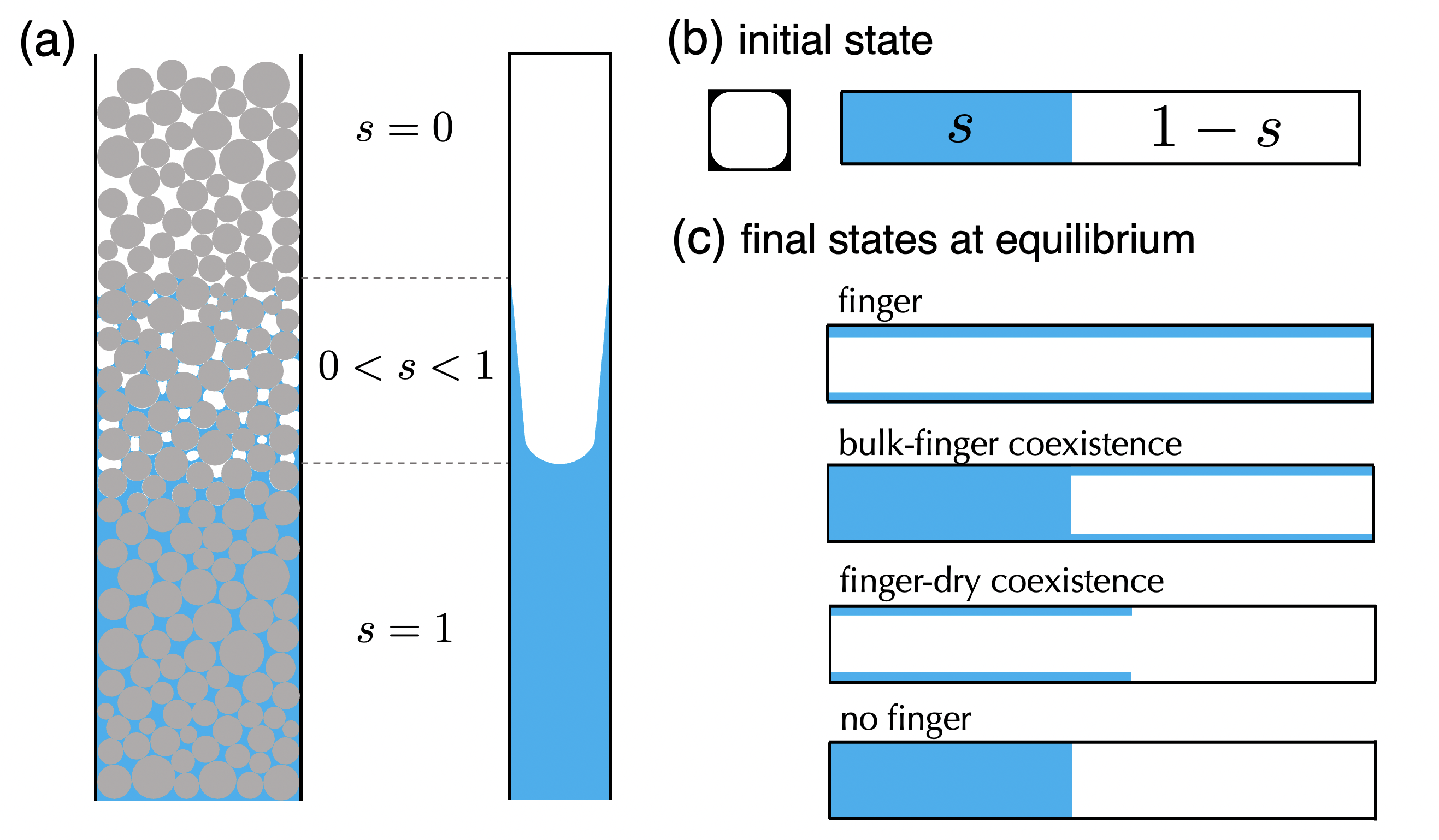}
  \caption{(a) Typical states in a porous medium and the corresponding states in a capillary tube. (b) Initial configuration. The liquid occupies a fraction $s$ in the tube. (c) Possible equilibrium configurations.}
  \label{fig:sat}
\end{figure}

In general, porous media exhibits complex pore structure, including broad distribution of pore size and varied junction geometry, which complicates comprehensive understanding of flow and transport phenomena \cite{Lehmann2008, Yiotis2012, Prat2007, Coussot2000}.
Simplified models have been used to understand the fluid transport in porous media.
An illustrative example is the study of evaporation dynamics, where the characteristic drying behavior of porous media is well represented by the liquid evaporation in a square tube \cite{Chauvet2009} (Fig.~\ref{fig:sat}(a)).
In the cornered tube, partially wet state corresponds to the fluid columns along the corners (also known as corner film, herein we shall call them ``fingers'' \cite{2018_square, 2021_rect_sstar, 2021_rect_LW, 2022_square_g, 2025_polygon}).

Here we study the wetting configuration in a square tube with rounded corners, with the aim to use it as a simple model for porous medium.  More specifically, we  address the following question:
\emph{In a tube filled by a fluid with volume fraction $0<s<1$, what is the equilibrium wetting configuration?}
The initial configuration is shown in Fig. \ref{fig:sat}(b), and  the possible final configurations (Fig. \ref{fig:sat}(c)) are: 
\begin{itemize}
\item partially wet state (finger)
\item finger coexisting with completely wet state (bulk-finger coexistence)
\item finger coexisting with dry state (finger-dry coexistence)
\item completely wet state coexisting with dry state (no finger)
\end{itemize}
It should be noted that here we are considering the states on the length scale much larger than the tube diameter (or the pore size in Fig. \ref{fig:sat}(a)), and ignore the detailed structure on the scale of tube diameter. 
In such treatment, the transition region between different wetting states is represented by a discontinuous jump in the saturation, shown schematically in Fig. \ref{fig:sat}(c).

\begin{figure}[ht]
  \includegraphics[width=0.6\columnwidth]{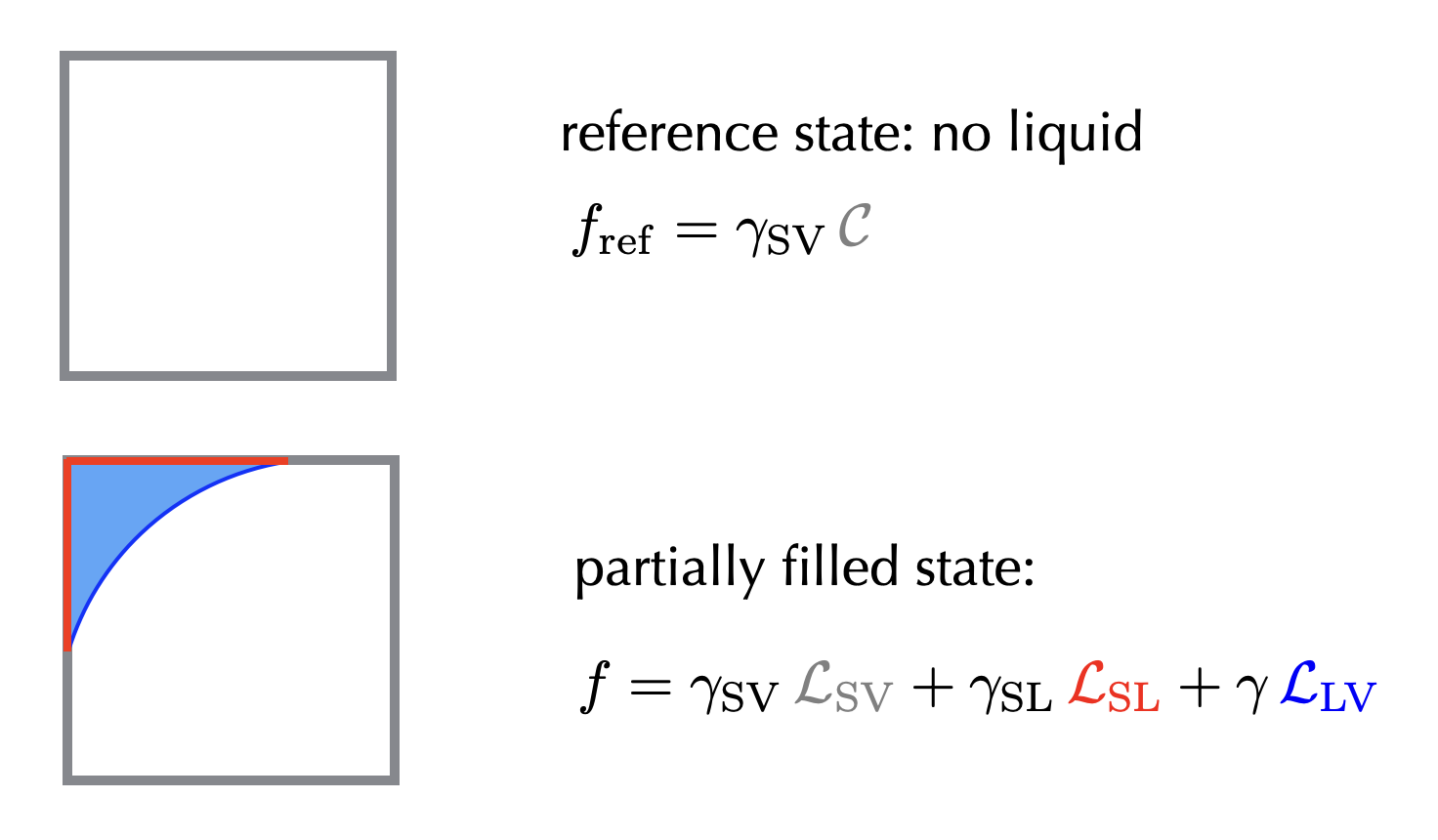}
  \caption{Calculation of the interfacial energy $f(s)$.}
  \label{fig:fs}
\end{figure}

Our approach is based on the wetting energy density function $f(s)$, which is  defined as 
the free energy per unit tube length when the saturation is $s$.  We choose the dry state as the reference state. The free energy of this state is $f_{\rm ref} = \gamma_{\rm SV} \mathcal{C}$, where $\mathcal{C}$ is the perimeter of the cross-section and $\gamma_{\rm SV}$ is the interfacial tension at the solid/vapor boundary (Fig. \ref{fig:fs}). In this paper the subscripts S, V, L stand for solid, vapor and liquid, respectively.
With this notation, the wetting energy $f(s)$ is generally written as
$f(s)=\gamma_{\rm SV} \mathcal{L}_{\rm SV} + \gamma_{\rm SL} \mathcal{L}_{\rm SL} + \gamma \mathcal{L}_{\rm LV} - \gamma_{\rm SV} \mathcal{C}$.
With the use of the geometric equality $\mathcal{C} = \mathcal{L}_{\rm SL} + \mathcal{L}_{\rm SV}$ and Young's equation $\gamma_{\rm SV} = \gamma_{\rm SL} + \gamma \cos\theta$ ($\theta$ being the equilibrium contact angle), $f(s)$ can be written as
\begin{equation}
	f(s) = - \gamma_{\rm SV} \mathcal{L}_{\rm SL} + \gamma_{\rm SL} \mathcal{L}_{\rm SL} + \gamma \mathcal{L}_{\rm LV} 
	=  (-\gamma \cos\theta) \mathcal{L}_{\rm SL} + \gamma \mathcal{L}_{\rm LV}  \, .
	\label{eq:fs} 
\end{equation}

Given $f(s)$, we can predict which state in Fig. \ref{fig:sat}(c) is realized at equilibrium.
Let $L$ be the total length of the tube, and $\phi_1$, $\phi_f$, and $\phi_0$ be the fractions of the completely wet state, partially wet state, and dry state  ($\phi_1 + \phi_f + \phi_0 = 1$).
The total energy is then given by
\begin{equation}
	\label{eq:Ftotal}
	F_{\rm total}/L = \phi_1 f(1) + \phi_f f(s) + \phi_0 f(0) \, .
\end{equation}
Here we have ignored the contribution of the transition regions where $s$ changes discontinuously. 
The equilibrium state is given by the state which minimizes $F_{\rm total} $ under the constraint that the fluid volume is fixed at $V_{\rm L} =( \phi_1 + \phi_f s)L S_0$ ($S_0$ being the area of the cross-section).

As an example, we consider the bulk-finger coexistence.
Since  there is no dry region, $\phi_0=0$ and $\phi_1 = 1 - \phi_f$. Therefore 
 $F_{\rm total}$ is written as
$F_{\rm total}/L = (1-\phi_f) f(1) + \phi_f f(s)$. The volume constraint can be accounted for 
by Lagrangian multiplier $\xi$, and the energy to be minimized is 
\begin{equation}
  \mathcal{F}/L 
    = \phi_f f(s) + (1-\phi_f) f(1) + \xi \Big[  (1-\phi_f) S_0 +  \phi_f S_0 s  - V_{\rm L}/L \Big] 
\end{equation}
This gives the following set of equations
\begin{align}
  \label{eq:dLds}
  \frac{\partial (\mathcal{F}/L)}{\partial s} &=  \phi_f \Big[ f'(s) + \xi S_0 \Big] = 0 \, ,\\
  \label{eq:dLdphi}
  \frac{\partial (\mathcal{F}/L)}{\partial \phi_f} &= f(s) - f(1) + \xi S_0 ( s-1 ) = 0 \, .
\end{align}
These equations determine the saturation $s^*$ of the partially wet state that coexists with the completely wet state:
\begin{equation}
	\label{eq:sstar}
	f'(s^*) = \frac{f(1)-f(s^*)}{1-s^*} \quad\quad  \text{ (bulk-finger coexistence) }
\end{equation}

The finger-dry coexistence can also be determined similarly, by setting $\phi_1 = 0$ and $\phi_0 = 1-\phi_f$.  This leads to the following condition
\begin{equation}
	\label{eq:sdstar}
	f'(s^{**}) = \frac{f(s^{**})}{s^{**}} \quad\quad\quad \text{(finger-dry coexistence)}
\end{equation}

The above argument is precisely the same as that used in the phase separation in binary mixtures.
In the binary mixture, the free energy is expressed as a function of solute concentration $c$, while in the present problem, the free energy is expressed as a function of liquid saturation $s$.
Since both concentration and saturation are conservative, we can use the same argument.
Furthermore, we can show that the system  is stable if $f''(s) > 0$, and unstable if $f''(s)<0$.  
In the case of binary mixtures, the first derivative $f'(c)$ corresponds to chemical potential.  
In the present system, the first derivative $f'(s)$ corresponds to the Laplace pressure $p_c$, which is the jump of the pressure across the liquid/vapor boundary.
The proof of this statement is given in Appendix \ref{app:pc}.
Following this analogy, we consider length scales much larger than the tube side-length $2a$. 
Transition region between coexisting phases contributes an energy that scales as $a/L$ relative to the bulk. 
For long tubes ($L \gg a$), this contribution is negligible.

Now we apply the above results to capillary tubes having some specific cross-sections. 

\begin{figure}[ht]
  \includegraphics[width=0.7\columnwidth]{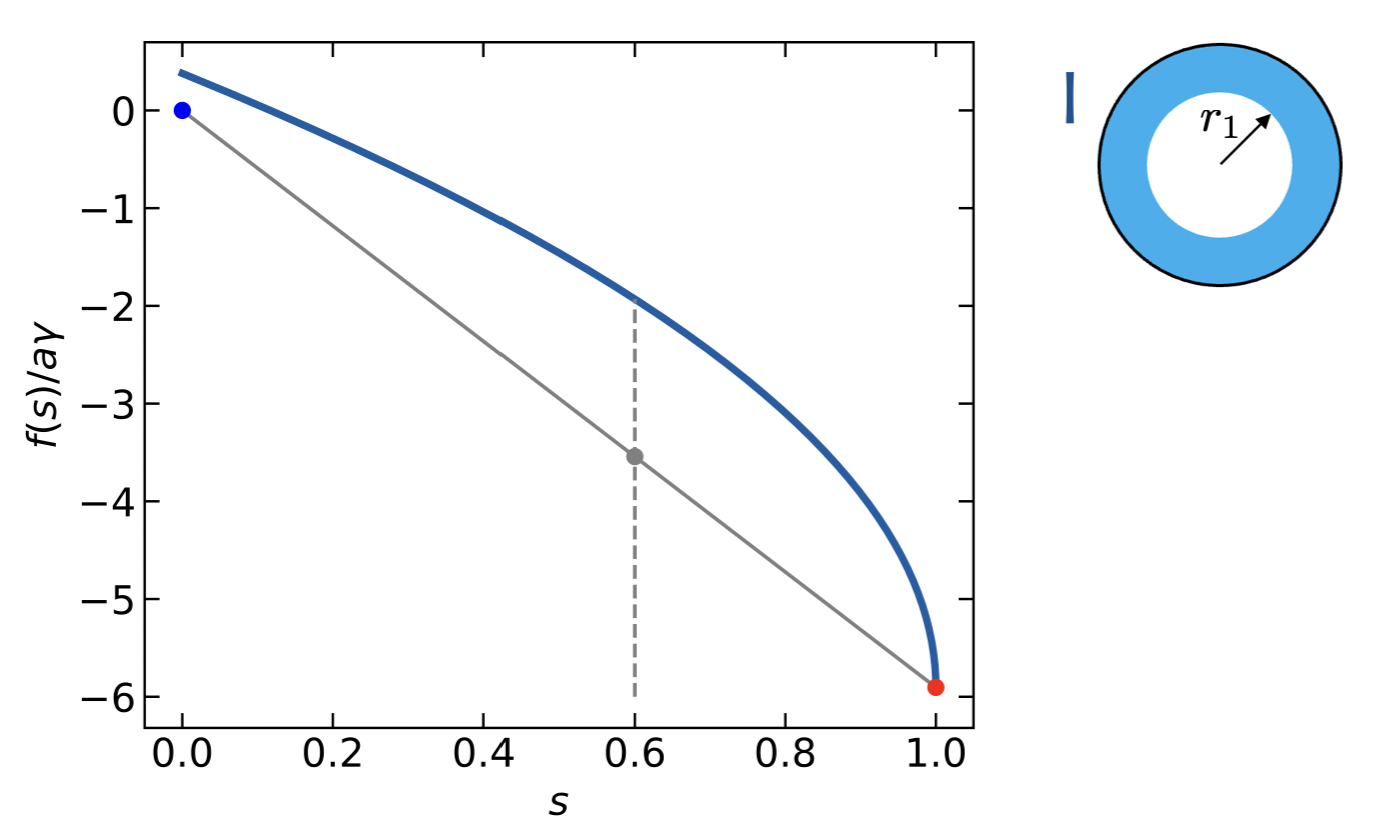}
  \caption{Interfacial energy $f(s)$ for a circular tube. The contact angle is $\theta=20^{\circ}$.}
  \label{fig:freeE_circle}
\end{figure}

\emph{Circular tube.} For a circular tube of radius $a$, the area of the cross-section is $S_0 = \pi a^2$. 
The partially filled case corresponds to a full circle of liquid-vapor interface with a radius $r_1 < a$ (Fig. \ref{fig:freeE_circle}I).
The saturation can be written a function of $r_1$, 
\begin{equation}
	s = 1- \frac{\pi r_1^2}{S_0} = 1 - \left( \frac{r_1}{a} \right)^2 \, .
\end{equation}
The interfacial energy $f(s)$ is expressed as 
\begin{equation}
	\label{eq:fE_circ}
	\frac{f(s)}{a \gamma}  = - 2 \pi \cos\theta + 2 \pi \sqrt{1-s} \, .
\end{equation}

This function is shown in blue in Fig. \ref{fig:freeE_circle} for $\theta=20^{\circ}$. 
The curve is upper convex for all saturation $s$, thus all partially wet states are unstable. 
If a state with saturation $s$ ($0<s<1$ ) splits into two states of saturation $s=0$ and $s=1$, each state occupies the fraction $1-s$ and $s$ of the tube length.
Therefore the energy becomes $(1-s)f(0) + s f(1)$  which is lower than the original energy $f(s)$ (see grey dot in Fig. \ref{fig:freeE_circle}). 
As a result, the finger will split into a dry state and a completely wet state.
The actual energy curve is given by the grey line  connecting the $s=0$ and $s=1$ points in Fig. \ref{fig:freeE_circle}.  This line corresponds to the convex envelope of the free 
energy (\ref{eq:fE_circ}) \cite{MaoSheng2019}.  

\begin{figure}[htbp]
  \includegraphics[width=1.0\columnwidth]{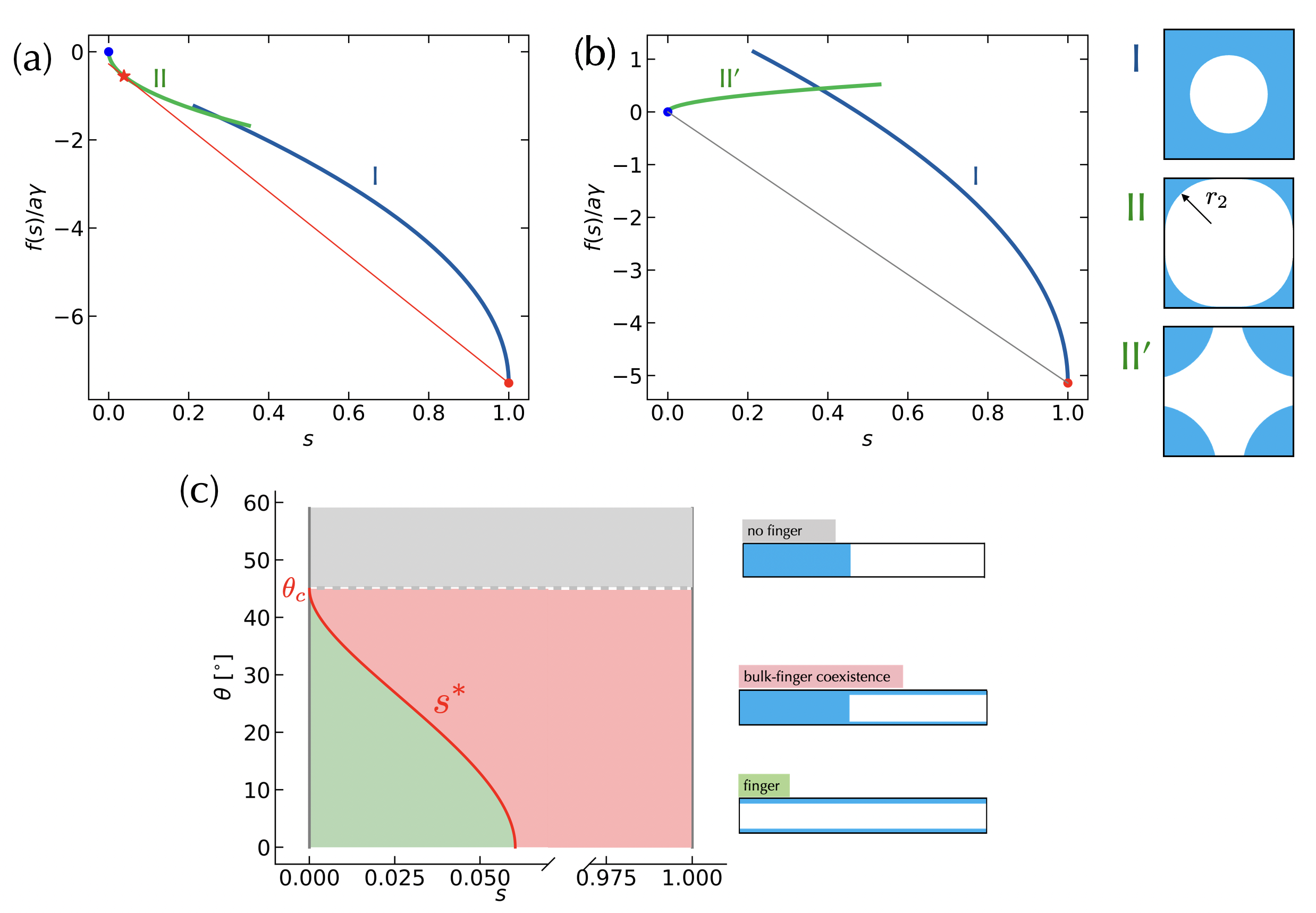}
  \caption{Interfacial energy $f(s)$ for a square tube. (a) Contact angle $\theta=20^{\circ}$. (b) $\theta=50^{\circ}$. Case I and II(II') are presented by the blue curves and green curves, respectively. The $s^*$ is given by the saturation at the red star. 
  (c) Phase diagram for a square tube.}
  \label{fig:square}
\end{figure}

\emph{Square tube} has corners and permits the corner film configuration as shown in  Fig. \ref{fig:square}II.
Let $r_2$ be the  radius of the curvature of the film surface.  The saturation $s$ is expressed  as a function of $r_2$ and the contact angle $\theta$ as 
\begin{equation}
	s 
	  = \left[ (\cos\theta - \sin\theta) \cos\theta - \left( \frac{\pi}{4} 
	    - \theta \right) \right] \left( \frac{r_2}{a} \right)^2  
 \equiv 
	 A(\theta) \left( \frac{r_2}{a} \right)^2 \, 
\end{equation}
The interfacial energy is expressed as
\begin{equation}
	\label{eq:fs_sII}
	\frac{f(s)}{a\gamma} = - {\rm sign}\left( \frac{\pi}{4} - \theta \right) 8 \sqrt{ A(\theta) } s^{1/2} \, .
\end{equation}
The energy is positive for convex surface ($\theta > 45^{\circ}$) and negative for concave surface ($\theta < 45^{\circ}$). 
The detailed derivation of these equations is given in  Appendix \ref{app:square}.

Figure \ref{fig:square}(a) shows $f(s)$ for $\theta=20^{\circ}$. 
The blue and green curves correspond to the case I and II, respectively. 
In case I, $f''(s)$ is negative and the uniform finger is unstable. 
In case II, $f''(s)$ is positive and the uniform finger becomes stable.  However,  this 
state is not the state of energy minimum.  By constructing the convex envelop of $f(s)$, one can show  that 
the uniform state satisfying the condition $s^*<s<1$ can lower the energy by separating into two states: the partially wet state of $s^*$ and the completely wet state of $s=1$. 

Figure \ref{fig:square}(b) shows $f(s)$ for $\theta=50^{\circ}$.  In this case, 
$f(s)$ is always above the line connecting $s=0$ and $s=1$ (the grey line in Fig. \ref{fig:square}(b)). 
Therefore, the system always separates into two states, the completely wet 
state of $s=1$, and dry state of $s=0$.  
In other word,  the fluid having contact angle  $\theta=50^{\circ}$ cannot take finger state, 
and behaves  as in the circular tube.  

The difference between the cases (a) and (b) can be made more clear in the phase diagram shown in
Figure \ref{fig:square}(c).  Here the three wetting states  in square tube are shown in the plane of 
the fluid contact angle $\theta$ and fluid volume fraction $s$.   It is seen that there is a critical contact angle
$\theta_c$.  If $\theta<\theta_c$, there is a characteristic saturation $s^*$ which is a function of $\theta$. If $s<s^*$,  the fluid forms a uniform finger  and if $s>s^*$, the fluid separates into two regions, one in partially wet state of $s^*$, and the other in completely wet state.
If $\theta > \theta_c$, the finger cannot exist and the fluid always take the bulk/dry coexistence.
The critical angle $\theta_c$ is $45^{\circ}$ and this is related to the Concus-Fin condition \cite{Concus1969} of the wetting of fluid in corners.

\begin{figure}[ht]
  \includegraphics[width=0.9\columnwidth]{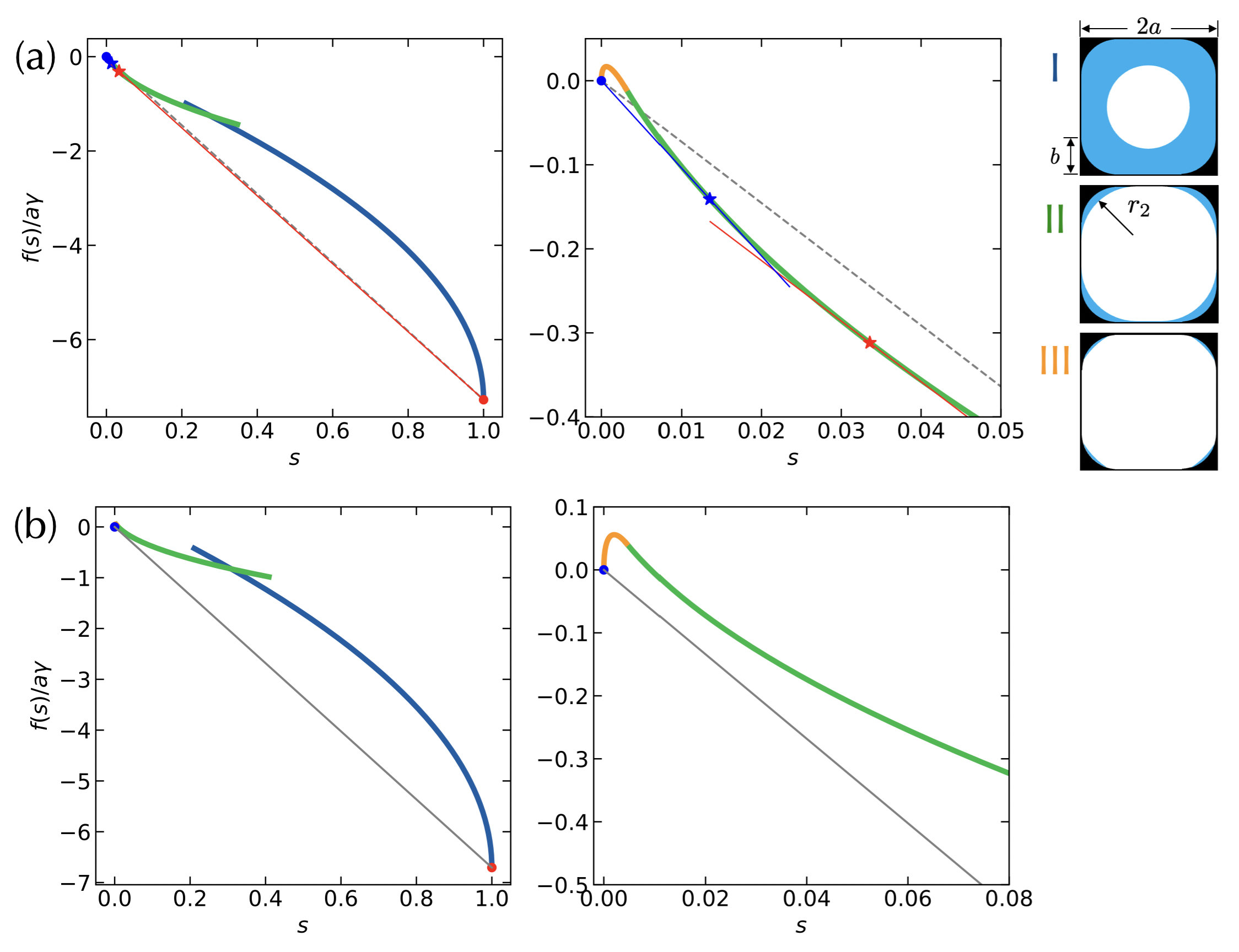}
  \includegraphics[width=0.6\columnwidth]{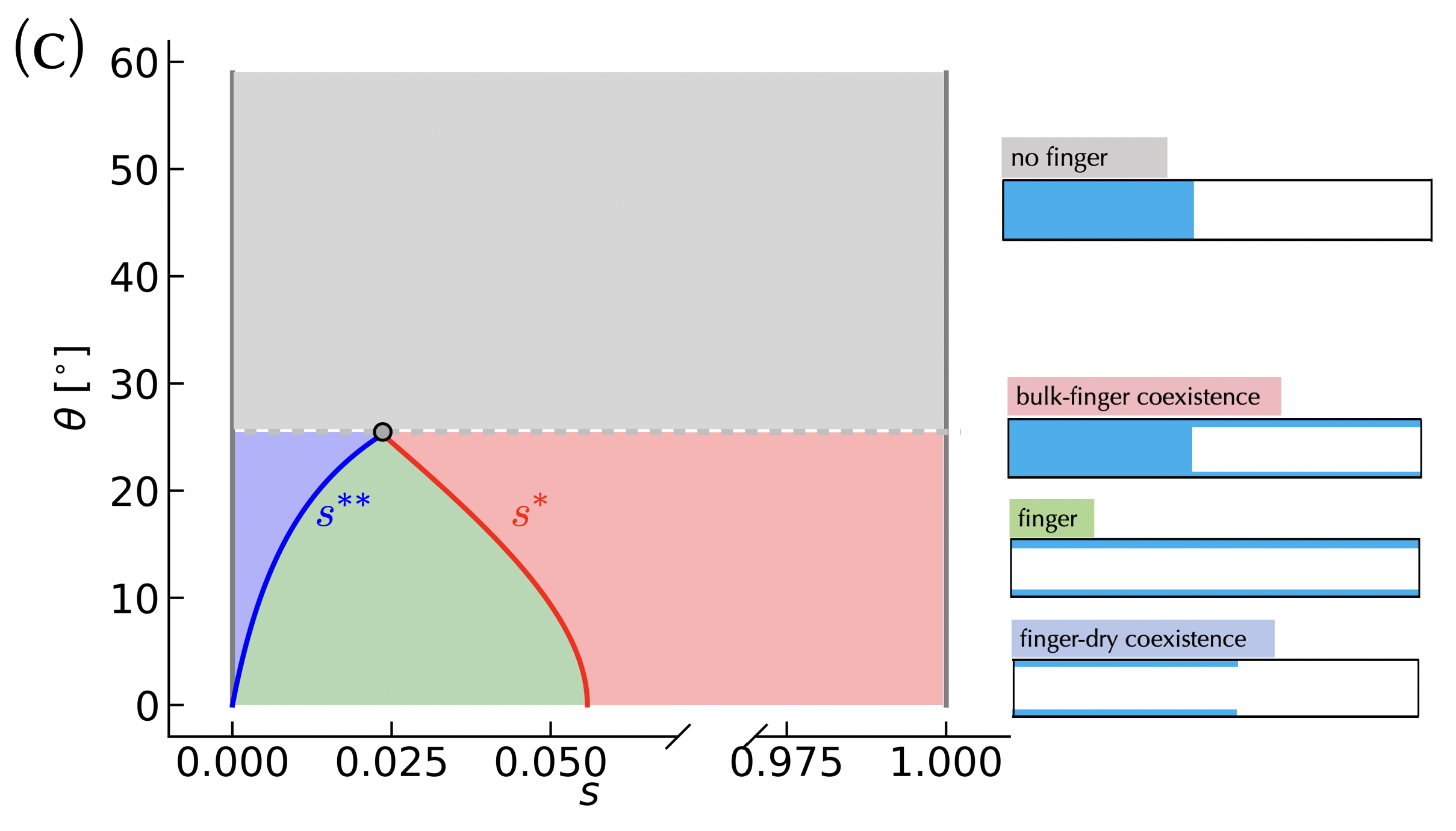}
  \caption{Interfacial energy $f(s)$ for a square tube with rounded corners. 
  (a) Contact angle $\theta=20^{\circ}$ and roundness $\beta=0.15$. 
  (b) Contact angle $\theta=30^{\circ}$ and roundness $\beta=0.15$. 
  Case I, II, III are presented by the blue, green, and orange curves, respectively. 
  $s^*$ and $s^{**}$ are denoted by the red and blue stars, respectively.
  The figures on the right are enlarged part of the curves where $s^*$ and $s^{**}$ are more apparent. 
  (c) Phase diagram for a square tube with rounded corners with $\beta=0.15$.} 
  \label{fig:rounded}
\end{figure}

\emph{Square tube with rounded corners.} We now consider the case that the corner of the square tube is not 
sharp, but is rounded, the radius of the curvature of which is $b$.
The roundness is represented by a dimensionless parameter $\beta=b/a$, where $2a$ is the side length of the square (see Fig. \ref{fig:rounded}I).  
As $\beta$ increases from 0 to 1, the cross-section of the tube changes smoothly from square to circle.

For the rounded corner, there are three cases shown in Fig. \ref{fig:rounded}. 
For each case, we can calculate the wetting energy $f(s)$ as a function of the saturation $s$  by geometrical consideration. (see Appendix \ref{app:rounded}).

Figure \ref{fig:rounded}(a) shows the plot of $f(s)$ for $\theta=20^{\circ}$ and $\beta=0.15$, where  $s^*$ and $s^{**}$ are denoted by the orange and the blue stars (see the enlarged view shown on the right).  For $s<s^{**}$, the partially wet state is unstable. The system separates into dry state of $s=0$ and partially wet state of $s=s^{**}$.  For $s^{**}<s<s^{*}$, the partially wet state is stable, and the tube is occupied by a uniform finger.
For $s>s^{*}$, the system again separates into the partially wet state of $s=s^*$ and the completely wet state of $s=1$ .

Figure \ref{fig:rounded}(b) shows the plot of $f(s)$ for $\theta=30^{\circ}$ and $\beta=0.15$. 
Here the energy curves lie above the line connecting $s=0$ and $s=1$ (the grey line). 
In this case, as in the case of circular tube and the case of $\theta>\theta_c$ in square tube,
the finger cannot exist and the fluid always take the bulk/dry coexistence.

For a fixed roundness parameter $\beta$, $s^*$ and $s^{**}$ are calculated as a function of the contact angle $\theta$ by Eqs. (\ref{eq:sstar}) and (\ref{eq:sdstar}).   
Figure \ref{fig:rounded}(c) shows the phase diagram obtained by such calculation for $\beta=0.15$.
There exists a critical contact angle where $s^*=s^{**}$. 
Beyond this angle, the finger regime disappears. 
The system then exhibits only the coexistence between the dry and completely wet states.

\begin{figure}[ht]
  \includegraphics[width=0.5\columnwidth]{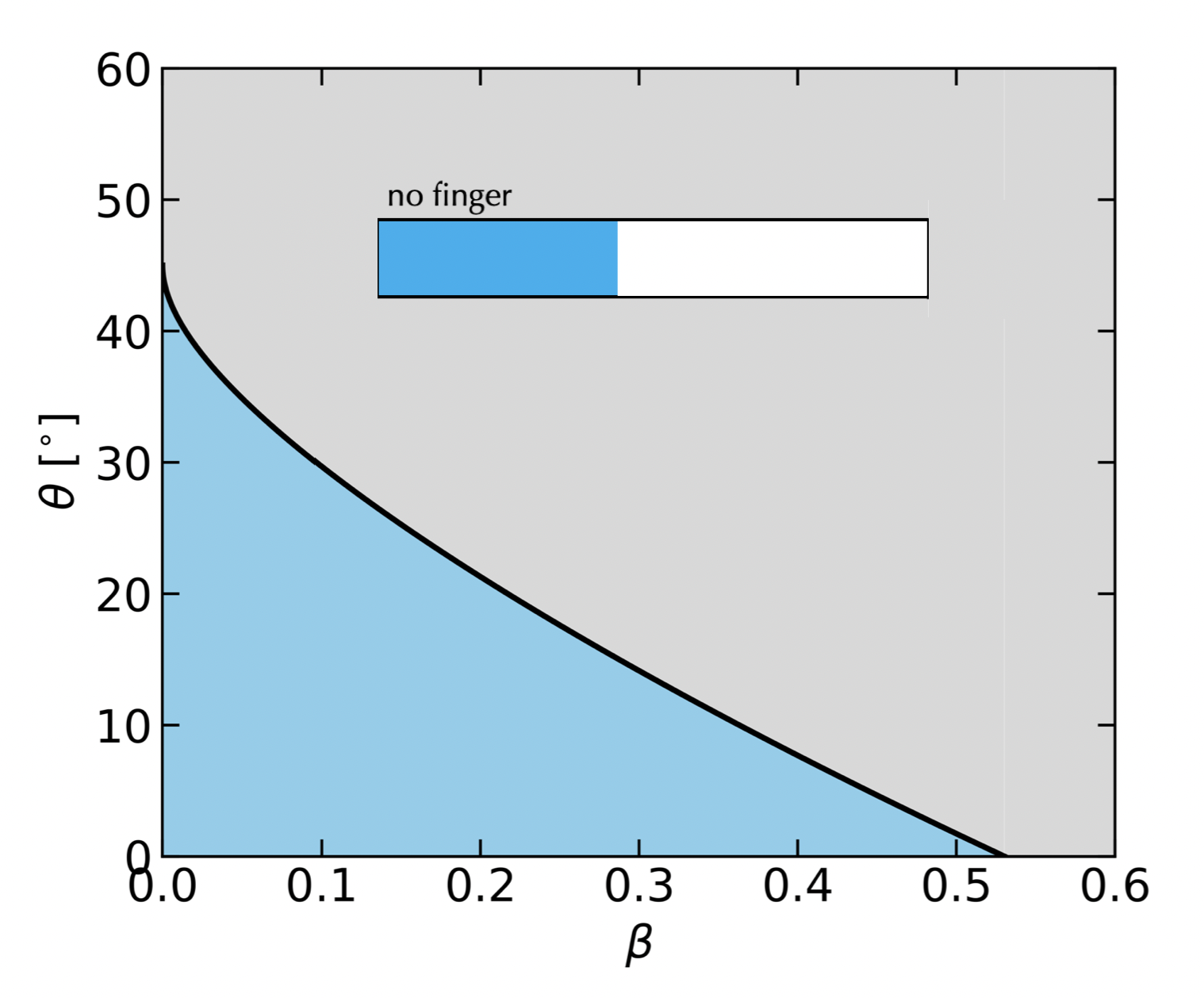}
  \caption{Phase diagram for a square tube with rounded corners in the plane of roundness and contact angle ($\beta$-$\theta$).  The boundary indicates the function $\theta_c(\beta)$ }
  \label{fig:pd_beta_theta}
\end{figure}

\emph{Existent condition for the finger.}  The critical angle $\theta_c$ is a function of the roundness parameter
$\beta$.  This gives a phase diagram in the $\beta - \theta$ plane (see Fig. \ref{fig:pd_beta_theta}). 
The finger cannot exist if $\theta > \theta_c(\beta)$; it  can exist only in the region of 
$\theta<\theta_c(\beta)$, either as a uniform state, or as the coexistent state with bulk or dry region.

For $\theta=0$, the critical roundness is calculated to be $\beta = (4 - 2 \sqrt{\pi})/(4- \pi) \simeq 0.53$ (Appendix \ref{app:pd}). 
This indicates that the fingers disappear before the cross-section becomes fully circular ($\beta=1$), a result consistent with prior calculation \cite{Concus1990}. 

\emph{Summary and Discussions.} In this work, we have developed a theory for  the wetting behavior of a fluid inside a capillary tube of uniform cross-section.  Our argument is the same as the  phase separation theory in binary mixtures, but we use the wetting energy density $f(s)$, which can be calculated for any tube having uniform cross section.
Our theory predicts the existence of three wetting states: a completely wet state, a partially wet state, where the fluid exists only as corner films or ``fingers'', and a completely dry state. 
From the shape of $f(s)$, coexistence among these three states can be determined. 
We have applied the theory to the case of a square tube with rounded corners.
We have shown that the partially wet state is possible only when certain condition is fulfilled.  

Our theoretical results have direct implications for applications involving porous media and microfluidic devices.
Corner flow plays a fundamental role in enhanced oil recovery and groundwater remediation. 
In these systems, the wetting phase resides in pore corners as films, enabling connected pathways and trapping mechanisms. 
Our theoretical approach is general and not limited to simple geometries. 
The only required input is $f(s)$, the free energy density as a function of saturation. 
For a given porous medium, $f(s)$ can be obtained experimentally by measuring capillary pressure at different saturations and integrating \cite{Bear}. 
Alternatively, one can compute $f(s)$ by evaluating the interfacial free energy for different wetting configurations at fixed saturation, using pore-scale models or image-based reconstructions of the pore space \cite{Cejas2017}. 
Once $f(s)$ is known, our method applies directly: analyzing its convexity and constructing common tangents. 
These analysis yield the equilibrium phase diagram, predicting which saturation states (fully wet, partially wet, or dry) are thermodynamically stable. 
This provides a predictive framework for determining when corner films remain connected, addressing a key uncertainty in current reservoir simulations.

The thermodynamic framework established here opens new avenues for exploration. 
The analogy with phase transitions suggests that wetting kinetics could be modelled using Cahn-Hilliard equations with $f(s)$ as the free energy functional, connecting pore-scale dynamics to the rich physics of phase ordering. 
Nucleation barriers for the formation of fingers or dry patches can also be studied within this framework.
For complex porous geometry, multiple local minima of $f(s)$ may exist. 
The convexity analysis provides a thermodynamic basis for understanding hysteresis induced by metastable states.

For evaporation in porous media, finger films sustain high evaporation rates by maintaining connected liquid pathways to the surface \cite{Chauvet2009}.
Our theory identifies the saturation $s^{**}$ at which fingers coexist with a fully dried region. 
When the local saturation falls below $s^{**}$, fingers depin from the surface and the evaporation rate drops sharply. 
This depinning might occur earlier than predicted by previous theories, which assumed fingers recede when its tip saturation approaches zero. 
The result has direct relevance to soil science, food processing, and construction materials, where the duration of high evaporation rates is critical.

Microfluidic devices also benefit from our analysis. 
In channels with non-circular cross-sections, finger films govern phenomena from capillary filling to solute transport. 
Competition between the bulk flow and finger flow determines imbibition rates in capillaries with corners \cite{2018_square, 2021_rect_LW, 2022_square_g, 2025_polygon, Kubochkin2022a}. 
Our phase diagram, quantifying how roundness $\beta$ and contact angle $\theta$ jointly control finger stability, offers design guidelines for lab-on-a-chip applications. 
Devices requiring rapid filling should maintain sharp corners (small $\beta$) and high wettability (small $\theta$). 
Conversely, to avoid unwanted film formation, one could intentionally round corners beyond the critical threshold where fingers disappear.

In summary, our results offer a quantitative phase diagram that can guide interpretation of experiments and inform engineering design across these applications. 
The key insight, that corner roundness not just contact angle fundamentally controls finger stability, provides a previously missing design parameter for controlling flow at the pore scale.

\begin{acknowledgments}
This research was supported by the Advanced Materials--National Science and Technology Major Project (2025ZD0614503), the National Natural Science Foundation of China (22373036) and R\&D Program of Guangzhou (2024D03J0007). The computation of this work was made possible by the facilities of Information and Network Engineering and Research Center of SCUT.
\end{acknowledgments}

\bibliography{wetting}

\newpage
\appendix
\setcounter{figure}{0}
\renewcommand{\thefigure}{S\arabic{figure}}

\section{Different wetting cases}

\begin{figure}[ht]
  \includegraphics[width=0.6\columnwidth]{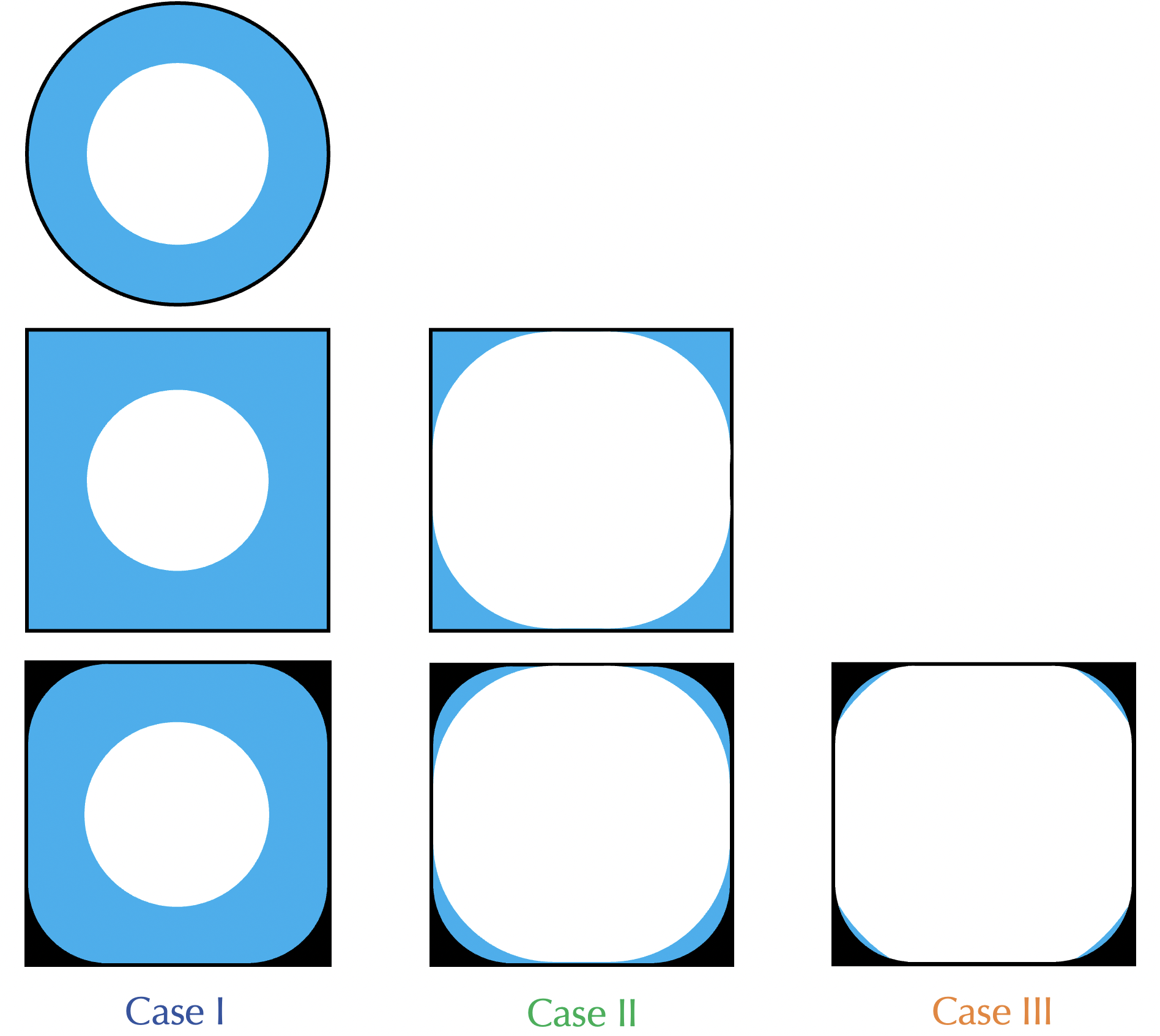}
  \caption{Wetting states in capillary tubes.}
  \label{fig:cases}
\end{figure}

Figure \ref{fig:cases} show different wetting states for a circular tube, a square tube, and a square tube with rounded corners (from top to bottom). 

\section{Connection to capillary pressure}
\label{app:pc}

Our results based on $f(s)$ are essentially the same as the conventional theory based on the Gauss equation of capillarity \cite{Hwang1977, Mason1984, Mayer1965, Princen1969, Princen1969a, Princen1970}.  When a (2d) fluid interface undergoes a variation of any kind, Gauss equation states there is a definite geometric relation between variation of solid-liquid interface $\ud\mathcal{L}_{\rm SL}$, variation of liquid-vapor interfaces $\ud\mathcal{L}_{\rm LV}$, and associated fluid area change $\ud S$
\begin{equation}
	\label{eq:Gauss}
	- k \, \ud S = - \cos\theta \, \ud\mathcal{L}_{\rm SL} + \ud\mathcal{L}_{\rm LV} \, ,
\end{equation}
where $k$ is the curvature of the fluid-vapor interface. 
Multiplying the above equation by the surface tension $\gamma$, we can identify the capillary pressure $p_c = \gamma k$ by Young-Laplace equation (here we have neglected the contribution from the curvature along the tube axis).
Comparing to Eq. (\ref{eq:fs}), the RHS of Eq. (\ref{eq:Gauss}) is just $\ud f(s)/\gamma$.
We then have
\begin{equation}
	\label{eq:pc}
	p_c = - \frac{\ud f(s)}{\ud S} =  - \frac{1}{S_0} \frac{\ud f(s)}{\ud s} = - \frac{f'(s)}{S_0} \, .
\end{equation}
The capillary pressure is proportional to the gradient of the energy function $f(s)$ with a negative sign. 
Since the liquid is driven from high $p_c$ region to low $p_c$ region, Eq. (\ref{eq:pc}) indicates that the state is unstable if the second derivative of $f(s)$ is negative, \emph{i.e.}, $f(s)$ is upper convex. 

\section{Square tube}
\label{app:square}

The detailed derivation of the interfacial energy in a square tube can be found in Ref. \cite{2018_square}.
Here we repeat the derivation for completeness, adopting the consistent notation of this manuscript.
The side length of the square cross-section is $2a$, and the resulting area is $S_0 = 4a^2$. 

\subsection{Case I: full-circle meniscus} 

Case I corresponds to a full circle of liquid-vapor interface.
The radius of the interface is $r_1$. 
The area of the liquid is $S = 4a^2 - \pi r_1^2$, thus the saturation is 
\begin{align}
	\label{eq:sat_sI}
	& s = \frac{S}{S_0} = 1 - \frac{\pi}{4} \left( \frac{r_1}{a} \right)^2 \,, \\
	\Rightarrow \quad & \frac{r_1}{a} = \sqrt{\frac{4}{\pi}} (1-s)^{1/2}  \,.
	\label{eq:r1_sI}
\end{align}

The maximum of $r_1$ is the half of the side-length, $r_1 \le a$. 
This leads to lower bound for the saturation $s_{\rm c1} = 1 - \pi/4$. 
The range of saturation in Case I is 
\begin{equation}
	s_{\rm c1} \le s \le 1, \quad s_{\rm c1} = 1 - \frac{\pi}{4} \, .
\end{equation}

The length of solid-liquid interface is $\mathcal{L}_{\rm SL} = 8a$, and the length of liquid-vapor interface is $\mathcal{L}_{\rm LV} = 2 \pi r_1$. 
The interfacial energy is then 
\begin{align}
	f(s) & = \mathcal{L}_{\rm SL} (-\gamma\cos\theta) + \mathcal{L}_{\rm LV} \gamma = - 8 a \gamma \cos\theta + 2\pi r_1 \gamma \\
	\frac{f(s)}{a\gamma} &= - 8 \cos\theta + 4 \sqrt{\pi} (1-s)^{1/2}
	\label{eq:freeE_sI}
\end{align}

\subsection{Case II, corner meniscus}

\begin{figure}[ht]
  \includegraphics[width=0.5\columnwidth]{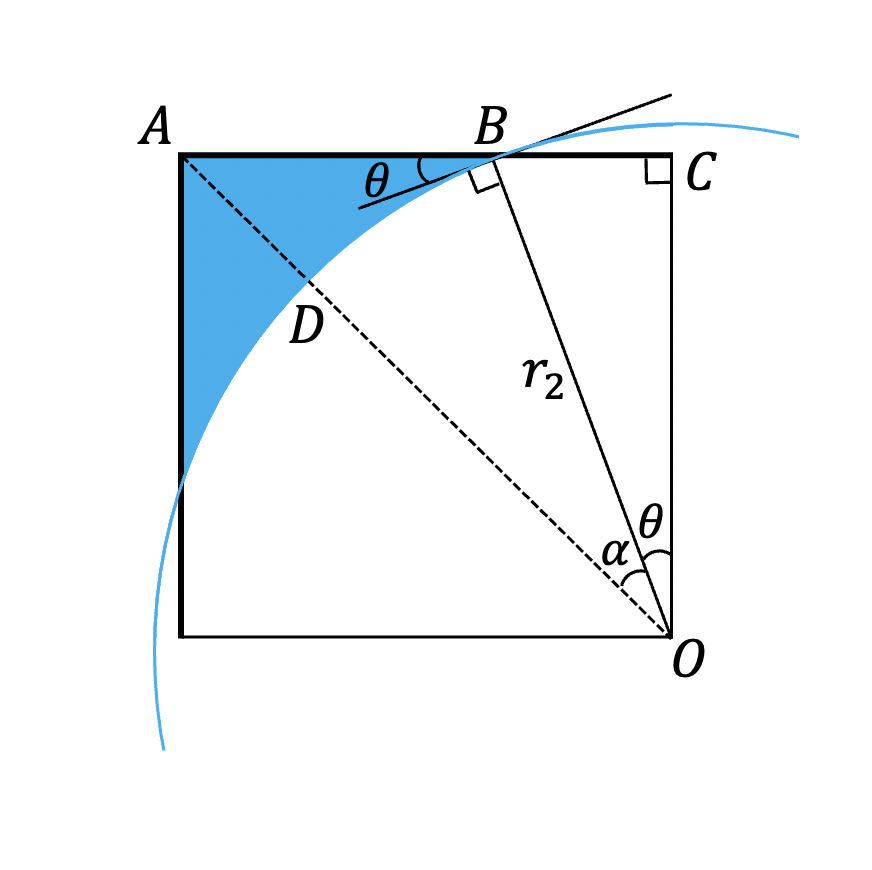}
  \caption{Square tube, case II with $\displaystyle \theta < \frac{\pi}{4}$. The finger meniscus is convex towards the corner.}
  \label{fig:square_caseII1}
\end{figure}

Case II corresponds to the fingers at four corners.
On one corner, the meniscus touches the side wall at point B with a contact angle $\theta$ (Fig. \ref{fig:square_caseII1}). 
This leads to $\angle{\rm BOC} = \theta$. 
One eighth of the meniscus $\overset{\frown}{\rm BD}$ is a portion of the circle with angle $\alpha = \frac{\pi}{4} - \theta$ and radius $r_2$. 

From the geometry, 
\begin{align}
	\overline{\rm OC} & = r_2 \cos\theta = \overline{\rm AC} \\
	\overline{\rm BC} & = r_2 \sin\theta \\
	\overline{\rm AB} & = \overline{\rm AC} - \overline{\rm BC} = r_2 (\cos\theta - \sin\theta) 
\end{align}
The area of $\rm{A}\overset{\frown}{\rm BD}$ is the area of triangle $\triangle\rm{ABO}$ subtracting the area of the sector $\rm{O}\overset{\frown}{\rm BD}$ 
\begin{align}
	S_{\triangle\rm{ABO}} & = \frac{1}{2} \overline{\rm AB} \times \overline{\rm OC} = \frac{1}{2} r_2^2 (\cos\theta - \sin\theta) \cos\theta \\
	S_{\rm{O}\overset{\frown}{\rm BD}} & = \frac{1}{2} \alpha r_2^2 = \frac{1}{2} \left( \frac{\pi}{4} - \theta \right) r_2^2 \\
	S_{\rm{A}\overset{\frown}{\rm BD}} & = S_{\triangle\rm{ABO}} - S_{\rm{O}\overset{\frown}{\rm BD}} = \frac{1}{2} r_2^2 \left[ (\cos\theta - \sin\theta) \cos\theta - \left(\frac{\pi}{4} - \theta \right) \right] 
\end{align}
The saturation is given by
\begin{align}
	\label{eq:sat_sII1}
	& s  = \frac{8 S_{\rm{A}\overset{\frown}{\rm BD}} }{S_0} = \left( \frac{r_2}{a} \right)^2 \left[ (\cos\theta - \sin\theta) \cos\theta - \left(\frac{\pi}{4} - \theta \right) \right] = \left( \frac{r_2}{a} \right)^2 A(\theta)\\
	\Rightarrow \quad & \frac{r_2}{a} = \frac{s^{1/2}}{\sqrt{A(\theta)}}
\end{align}
Here we define a function $A(\theta)$ to simplify the notation
\begin{equation}
	\label{eq:A}
	A(\theta) = (\cos\theta - \sin\theta) \cos\theta - \left( \frac{\pi}{4} - \theta \right) \, .
\end{equation}

The maximum of $\overline{\rm AB}$ is the half of the side-length, $\overline{\rm AB} \le a$. 
This leads to an upper bound for the radius $r_2/a < (\cos\theta - \sin\theta)^{-1}$. 
The range of saturation in Case II is 
\begin{equation}
	0 \le s \le s_{\rm c2} , \quad s_{\rm c2} = \frac{ A(\theta) }{ (\cos\theta - \sin\theta)^2} \, .
\end{equation}

The length of solid-liquid interface is $\mathcal{L}_{\rm SL} = 8 \times \overline{\rm AB}$, and the length of liquid-vapor interface is $\mathcal{L}_{\rm LV} = 8 \times \overset{\frown}{\rm BD} = 8 \alpha r_2$. 
The interfacial energy is then 
\begin{align}
	f(s) & = \mathcal{L}_{\rm SL} (-\gamma\cos\theta) + \mathcal{L}_{\rm LV} \gamma \nonumber \\
	& = 8  \times \overline{\rm AB} \times (-\gamma \cos\theta) + 8 \left( \frac{\pi}{4} - \theta \right) r_2 \gamma \\
	\frac{f(s)}{a\gamma} &= - 8 \frac{r_2}{a} A(\theta) = - 8 \sqrt{A(\theta)} s^{1/2}
	\label{eq:fs_sII1}
\end{align}

\begin{figure}[ht]
  \includegraphics[width=0.5\columnwidth]{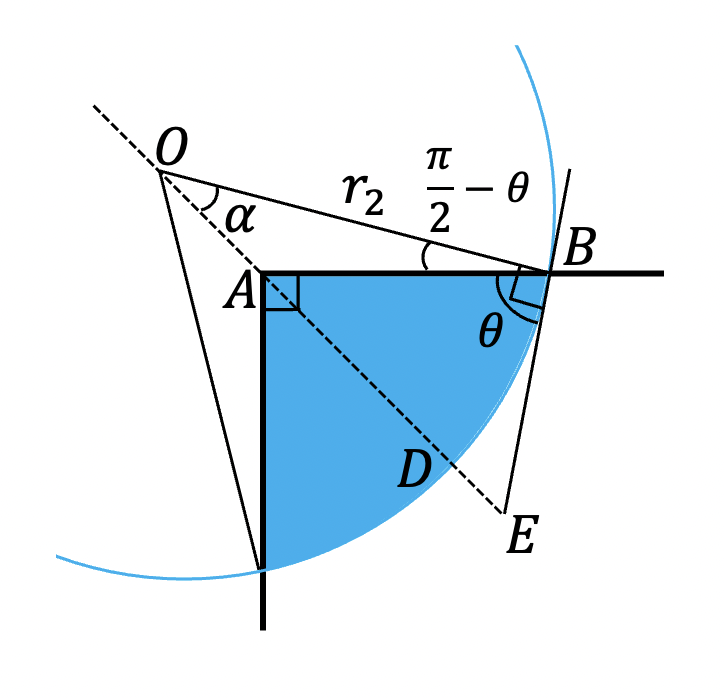}
  \caption{Square tube, case II with $\displaystyle \theta > \frac{\pi}{4}$. The finger meniscus is convex towards the tube center.}
  \label{fig:square_caseII2}
\end{figure}

The above derivation only applies for $\displaystyle \theta < \frac{\pi}{4}$. 
When $\displaystyle \theta > \frac{\pi}{4}$, the meniscus is convex towards the tube center (Fig. \ref{fig:square_caseII2}). 

The area of sector $\rm{O}\overset{\frown}{\rm BD}$ is 
\begin{equation}
	S_{\rm{O}\overset{\frown}{\rm BD}} = \frac{1}{2} \alpha r_2^2 = \frac{1}{2} \left( \theta - \frac{\pi}{4} \right) r_2^2 \, .
\end{equation}
The length of $\overline{\rm AB}$ is
\begin{equation}
	\overline{\rm AB} = r_2 (\sin\theta - \cos\theta) \, .
\end{equation}
The area of triangle $\triangle{\rm OAB}$ is 
\begin{equation}
	S_{\triangle{\rm OAB}} = \frac{1}{2} \overline{AB} \times r_2 \sin\left( \frac{\pi}{2} - \theta \right) = \frac{1}{2} r_2^2 (\sin\theta - \cos\theta) \cos\theta \, .
\end{equation}
The area of $\rm{A}\overset{\frown}{\rm BD}$ is $S_{\rm{A}\overset{\frown}{\rm BD}} = S_{\rm{O}\overset{\frown}{\rm BD}} - S_{\triangle{\rm OAB}}$. 
This leads to the saturation 
\begin{align}
	\label{eq:sat_sII2}
	& s = \frac{8 S_{\rm{A} \overset{\frown}{\rm BD}}}{S_0} = \left( \frac{r_2}{a} \right)^2 \left[ \left( \theta - \frac{\pi}{4} \right) - (\sin\theta - \cos\theta) \cos\theta \right] = \left( \frac{r_2}{a} \right)^2 A(\theta) \, \\
	\Rightarrow \quad & \frac{r_2}{a} = \frac{s^{1/2}}{\sqrt{A(\theta)}} \, .
\end{align}
Equation (\ref{eq:sat_sII2}) for $\displaystyle \theta > \frac{\pi}{4}$ has the same form of Eq. (\ref{eq:sat_sII1}) for $\displaystyle \theta < \frac{\pi}{4}$. 

The interfacial energy is given by
\begin{align}
	f(s) & = - 8  \times \overline{\rm AB} \times \gamma \cos\theta + 8 \alpha r_2 \gamma \\
	\frac{f(s)}{a\gamma} &= 8 \frac{r_2}{a} A(\theta) = 8 \sqrt{A(\theta)} s^{1/2} 
	\label{eq:fs_sII2}
\end{align}
Equation (\ref{eq:fs_sII2}) differs from Eq. (\ref{eq:fs_sII1}) only by a minus sign.

\subsection{Summary of square tube $f(s)$}

We summarize the interfacial energy functions for square tube.
\begin{itemize}

\item {\bf case I}: $s_{\rm c1} \le s \le 1$,  
$\displaystyle s_{\rm c1} = 1 - \frac{\pi}{4}$
\begin{equation}
	\frac{f(s)}{a\gamma} = - 8 \cos\theta + 4 \sqrt{\pi} (1-s)^{1/2}
\end{equation}

\item {\bf case II}: $0 \le s \le s_{\rm c2}$, 
$\displaystyle s_{\rm c2} = \frac{ A(\theta) }{ (\cos\theta - \sin\theta)^2}$
\begin{equation}
	\frac{f(s)}{a\gamma} = - {\rm sign} \left( \frac{\pi}{4}-\theta \right) 8 \sqrt{A(\theta)} s^{1/2} 
\end{equation}
$A(\theta)$ is given in Eq. (\ref{eq:A}). 

\end{itemize}

\subsection{Analysis of interfacial energy $f(s)$}

\begin{figure}[ht]
  \includegraphics[width=1.0\columnwidth]{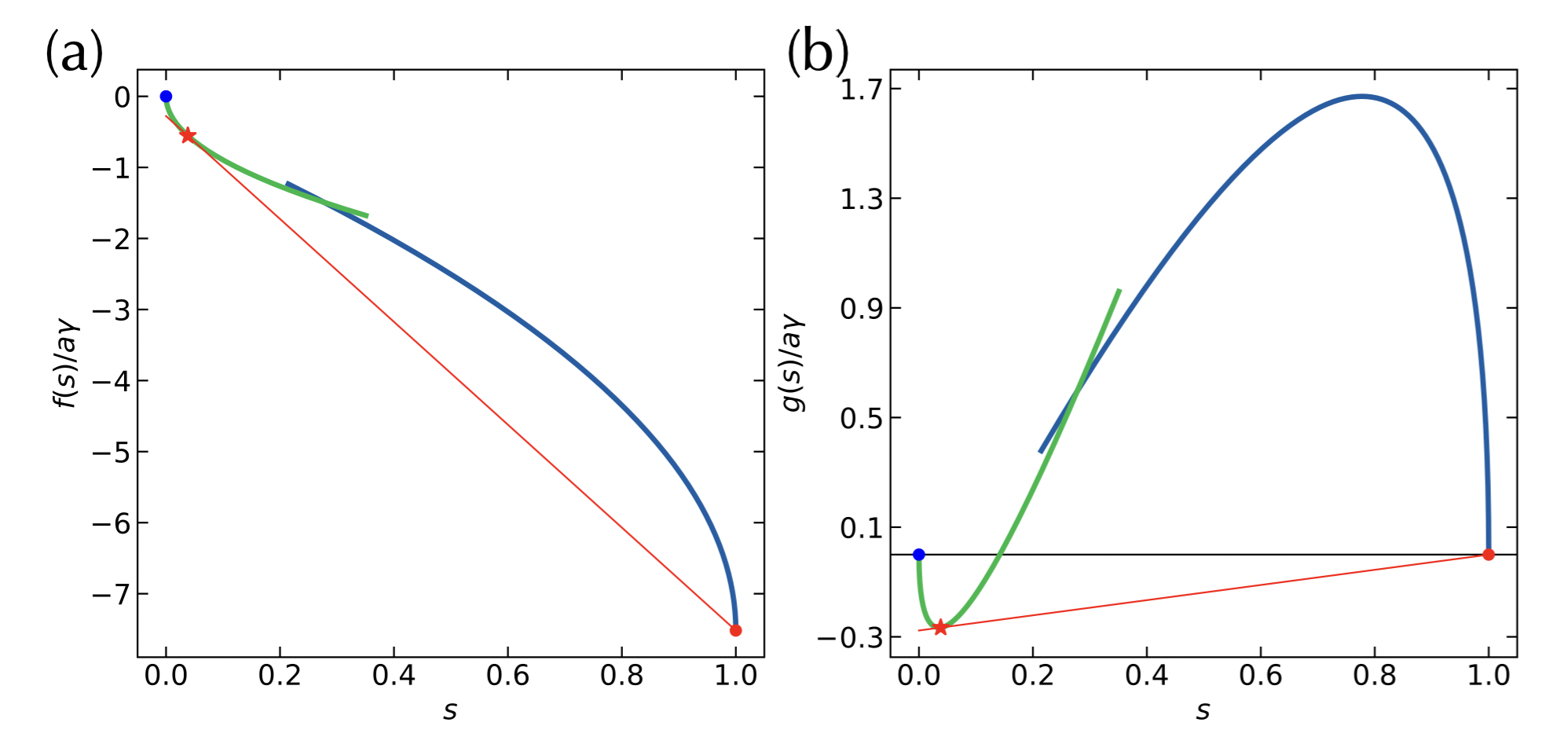}
  \caption{Interfacial energy curves $f(s)$ and $g(s)$ for a square tube. The contact angle $\theta = 20^{\circ}$.}
  \label{fig:freeE_square_app1}
\end{figure}

The interfacial energy curves for $\theta=20^{\circ}$ is shown in Fig. \ref{fig:freeE_square_app1}(a).
The value of $f(s)$ varies significantly between $s=0$ and $s=1$, making the line of tangency difficult to distinguish.
For clear illustration, we define a new function $g(s)$
\begin{equation}
	g(s) = f(s) - f(1) \times s \, .
\end{equation}
This function satisfies $g(0)=0$ and $g(1)=0$.
It quantifies the deviation from the straight line connecting the $s=0$ and $s=1$ endpoints.

\begin{figure}[ht]
  \includegraphics[width=1.0\columnwidth]{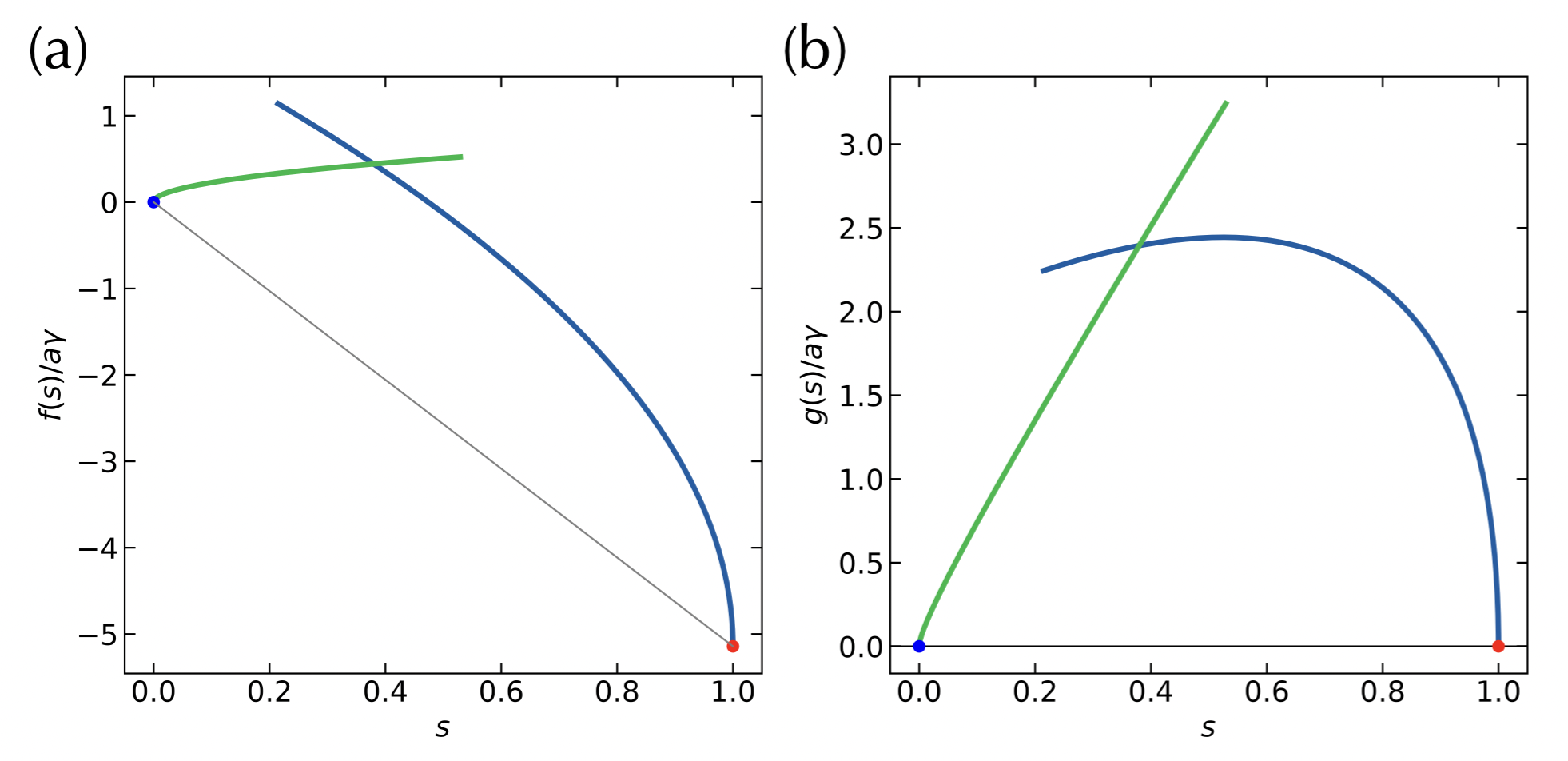}
  \caption{Interfacial energy curves $f(s)$ and $g(s)$ for a square tube. The contact angle $\theta = 50^{\circ}$.}
  \label{fig:freeE_square_app2}
\end{figure}

The gradient of the interfacial energy (for $\displaystyle \theta < \frac{\pi}{4}$) is
\begin{equation}
	\frac{f'(s)}{a\gamma} = - 4 \sqrt{ A(\theta) } s^{-1/2}
\end{equation}
We can identify the value of $s^*$ by 
\begin{align}
	& f'(s) = \frac{f(1) - f(s)}{1-s} \, \nonumber \\
	\Rightarrow \quad & \sqrt{A(\theta)} s - 2 \cos\theta s^{1/2} + \sqrt{A(\theta)} = 0
\end{align}
The solution is
\begin{equation}
	s^* = \frac{ \cos\theta - \sqrt{ \cos^2\theta - A(\theta)} }{ \sqrt{A(\theta)} }
\end{equation}
The solution of $s^*$ exists only when $\cos^2 \theta > A(\theta)$. 
Note that 
\begin{equation}
	A(\theta) = \cos^2\theta - \cos\theta \sin\theta - \left( \frac{\pi}{4} - \theta \right),
\end{equation}
Since the last two terms are positive for $\displaystyle \theta < \frac{\pi}{4}$, we always have a solution $s^*$ for $\displaystyle \theta < \frac{\pi}{4}$. 
When $\displaystyle \theta \rightarrow \frac{\pi}{4}$, the saturation $s^* \rightarrow 0$. 

The interfacial energy curves for $\theta=50^{\circ}$ is shown in Fig. \ref{fig:freeE_square_app2}.
In this case, one cannot find a solution for $s^*$, \emph{i.e.}, there is no fingers. 
The existent condition for the finger in a square tube is $\displaystyle \theta < \frac{\pi}{4}$.
The phase diagram is shown in Fig. \ref{fig:square}(c).

\section{Square tube with rounded corners}
\label{app:rounded}

The side length of the square is $2a$, making the area of the square $4a^2$. 
For rounded corner, the sharp normal angle is replace by a quarter of the circular arc of radius $b$. 
The roundness parameter $\beta$ is defined by $\beta = b/a$. 
The total area of the hollow section is 
\begin{equation}
  S_0 = 4a^2 - 4b^2 \left( 1 - \frac{\pi}{4} \right) = 4a^2 \left( 1 - \beta^2 \left( 1 - \frac{\pi}{4} \right) \right) = 4 a^2 B_2(\beta) \,.
\end{equation}
Here we introduce two functions of $\beta$ for notation simplification.
\begin{align}
	\label{eq:B1}
	B_1(\beta) & = 1 - \beta \left( 1 - \frac{\pi}{4} \right) \,, \\
	B_2(\beta) & = 1 - \beta^2 \left( 1 - \frac{\pi}{4} \right) \,.
	\label{eq:B2}
\end{align}
$B_1(\beta)$ will be applied later. 
When $\beta$ changes from 0 to 1, we have a smooth transition from a square tube (with side length $2a$) to a circular tube (with diameter $2a$).

\subsection{Case I: full-circle meniscus}

Case I corresponds to a full circle of liquid-vapor interface of radius $r_1$. 
The area of the liquid is given $S=S_0 - \pi r_1^2$. 
The saturation is 
\begin{align}
	& s = \frac{S}{S_0} = 1 - \frac{\pi}{4 B_2(\beta)} \left( \frac{r_1}{a} \right)^2 \,, \\
	\Rightarrow \quad & \frac{r_1}{a} = \sqrt{\frac{4B_2(\beta)}{\pi}} (1-s)^{1/2}  \,.
\end{align}
The range of saturation in Case I is 
\begin{equation}
	s_{\rm c1} \le s \le 1, \quad s_{\rm c1} = 1 - \frac{\pi}{4 B_2(\beta)} \, .
\end{equation}

The length of solid-liquid interface is $\mathcal{L}_{\rm SL} = 8a-8b+2\pi b = 8 a B_1(\beta)$, and the length of liquid-vapor interface is $\mathcal{L}_{\rm LV} = 2 \pi r_1$. 
The interfacial energy is then 
\begin{align}
	f(s) & = \mathcal{L}_{\rm SL} (- \gamma \cos\theta) + \mathcal{L}_{\rm LV} \gamma = - 8 a \gamma \cos\theta B_1(\beta) + 2\pi r_1^2 \gamma \\
	\frac{f(s)}{a\gamma} &= - 8 \cos\theta B_1(\beta) + 4 \sqrt{\pi B_2(\beta)} (1-s)^{1/2}
\end{align}
For sharp corners $\beta=0$, we have $B_1(\beta=0)=1$ and $B_2(\beta=0)=1$.
This recovers the expression in Eq. (\ref{eq:freeE_sI}) for a square tube.

\subsection{Case II: corner meniscus, contact line on the straight side}

\begin{figure}[ht]
  \includegraphics[width=0.5\columnwidth]{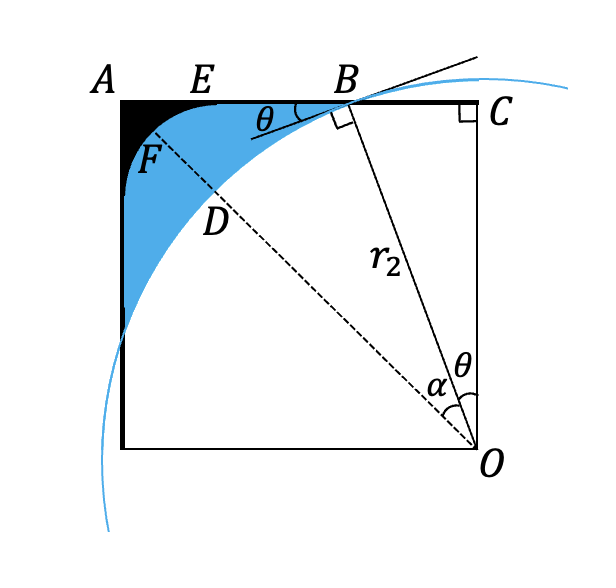}
  \caption{Square tube with rounded corners, case II with $\displaystyle \theta < \frac{\pi}{4}$. The finger meniscus is convex towards the corner.}
  \label{fig:rounded_caseII1}
\end{figure}

Case II of rounded corner is similar to the sharp corner. 
On one corner, the meniscus touches the side wall at point B with a contact angle $\theta$. 
This leads to $\angle{\rm BOC} = \theta$ (Fig. \ref{fig:rounded_caseII1}).
One eighth of the meniscus $\overset{\frown}{\rm BD}$ is a portion of the circle with angle $\alpha = \frac{\pi}{4} - \theta$ and radius $r_2$. 

From the geometry, 
\begin{align}
	\overline{\rm OC} & = r_2 \cos\theta = \overline{\rm AC} \\
	\overline{\rm BC} & = r_2 \sin\theta \\
	\overline{\rm AB} & = \overline{\rm AC} - \overline{\rm BC} = r_2 (\cos\theta - \sin\theta) 
\end{align}
We calculate the area of meniscus $\overset{\frown}{\rm BD} \overset{\frown}{\rm FE}$ by subtracting the sector $\rm{O}\overset{\frown}{\rm BD}$ and the rounded corner  $\rm{A}\overset{\frown}{\rm EF}$, from the triangle $\triangle\rm{ABO}$
\begin{align}
	S_{\triangle\rm{ABO}} & = \frac{1}{2} \overline{\rm AB} \times \overline{\rm OC} = \frac{1}{2} r_2^2 (\cos\theta - \sin\theta) \cos\theta \\
	S_{\rm{O}\overset{\frown}{\rm BD}} & = \frac{1}{2} \alpha r_2^2 = \frac{1}{2}  r_2^2 \left( \frac{\pi}{4} - \theta \right) \\
	S_{\rm{A}\overset{\frown}{\rm EF}} & = \frac{1}{2} b^2 \left( 1 - \frac{\pi}{4} \right) \\
	S_{\overset{\frown}{\rm BD} \overset{\frown}{\rm FE}} & = S_{\triangle\rm{ABO}} - S_{\rm{O}\overset{\frown}{\rm BD}} - S_{\rm{A}\overset{\frown}{\rm EF}} \nonumber \\
	& = \frac{1}{2} r_2^2 \left[ (\cos\theta - \sin\theta) \cos\theta - \left(\frac{\pi}{4} - \theta \right) \right] - \frac{1}{2} b^2 \left( 1 - \frac{\pi}{4} \right)
\end{align}
The saturation is given by
\begin{align}
	\label{eq:sat_rII1}
	& s  = \frac{8 S_{\overset{\frown}{\rm BD} \overset{\frown}{\rm FE}}  }{S_0} = \left( \frac{r_2}{a} \right)^2 \frac{ A(\theta) }{B_2(\beta)}  - \frac{1-B_2(\beta) }{ B_2(\beta) } \\
	\Rightarrow \quad & \frac{r_2}{a} = \frac{\sqrt{ B_2(\beta) s + (1-B_2(\beta)) }}{\sqrt{A(\theta)}}
\end{align}
where $A(\theta)$ and $B_2(\beta)$ are defined in Eqs. (\ref{eq:A}) and (\ref{eq:B2}).

The maximum of $\overline{\rm AB}$ is the half of the side-length $a$. 
This leads to an upper bound for the saturation
\begin{equation}
	\label{eq:sc2}
	s_{\rm c2} = \frac{ A(\theta) (\cos\theta - \sin\theta)^{-2} - (1-B_2(\beta)) }{ B_2(\beta) }
\end{equation}
The minimum of $\overline{\rm AB}$ is $b$. 
This leads to a lower bound for the saturation
\begin{equation}
	\label{eq:sc3}
	s_{\rm c3} = \frac{ \beta^2 A(\theta) (\cos\theta - \sin\theta)^{-2} - (1-B_2(\beta)) }{ B_2(\beta) }
\end{equation}
The range of saturation in case II is 
\begin{equation}
	s_{\rm c3} < s < s_{\rm c2} \, .
\end{equation}

The length of solid-liquid interface is $\mathcal{L}_{\rm SL} = 8 \times \overline{\rm AB} - 8b + 2\pi b$, and the length of liquid-vapor interface is $\mathcal{L}_{\rm LV} = 8 \times \overset{\frown}{\rm BD} = 8 \alpha r_2$. 
The interfacial energy is
\begin{align}
	f(s) & = \mathcal{L}_{\rm SL} (-\gamma\cos\theta) + \mathcal{L}_{\rm LV} \gamma \nonumber \\
	& = 8  \times \left( \overline{\rm AB} - b \left(1-\frac{\pi}{4} \right) \right) \times (-\gamma \cos\theta) + 8 \left( \frac{\pi}{4} - \theta \right) r_2 \gamma \\
	\frac{f(s)}{a\gamma} &= 8(1-B_1(\beta)) \cos\theta - 8 \sqrt{A(\theta)} \sqrt{ B_2(\beta) s + (1-B_2(\beta)) }
	\label{eq:fs_rII1}
\end{align}
This reduces to Eq. (\ref{eq:fs_sII1}) for $\beta=0$.

\begin{figure}[ht]
  \includegraphics[width=0.5\columnwidth]{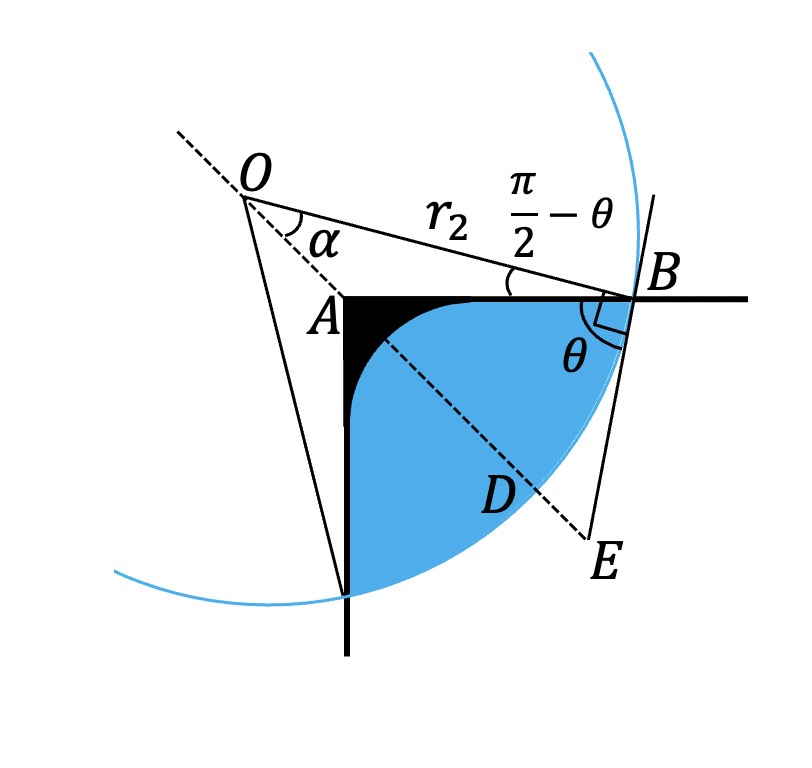}
  \caption{Square tube with rounded corners, case II with $\displaystyle \theta > \frac{\pi}{4}$. The finger meniscus is convex towards the tube center.}
  \label{fig:rounded_caseII2}
\end{figure}

The above derivation only applies for $\displaystyle \theta < \frac{\pi}{4}$. 
For the opposite case of $\displaystyle \theta > \frac{\pi}{4}$ (Fig. \ref{fig:rounded_caseII2}), the meniscus is convex towards the tube center. 

The saturation is
\begin{align}
	\label{eq:sat_rII2}
	& s = \left( \frac{r_2}{a} \right)^2 \frac{A(\theta)}{B_2(\beta)} - \frac{1-B_2(\beta)}{B_2(\beta)} \\
	\Rightarrow \quad & \frac{r_2}{a} = \frac{ \sqrt{ B_2(\beta) s + (1-B_2(\beta)) }}{ \sqrt{A(\theta)} } \, .
\end{align}
Equation (\ref{eq:sat_rII2}) for $\displaystyle \theta > \frac{\pi}{4}$ has the same form of Eq. (\ref{eq:sat_rII1}) for $\displaystyle \theta < \frac{\pi}{4}$. 

The interfacial energy is 
\begin{equation}
	\frac{f(s)}{a\gamma} = 8(1-B_1(\beta)) \cos\theta + 8 \sqrt{A(\theta)} \sqrt{ B_2(\beta) s + (1-B_2(\beta)) }
	\label{eq:fs_rII2}
\end{equation}
Equation (\ref{eq:fs_rII2}) differs from Eq. (\ref{eq:fs_rII1}) only by a sign in the second term.

\subsection{Case III: corner meniscus, contact line on the curved side}

\begin{figure}[ht]
  \includegraphics[width=0.5\columnwidth]{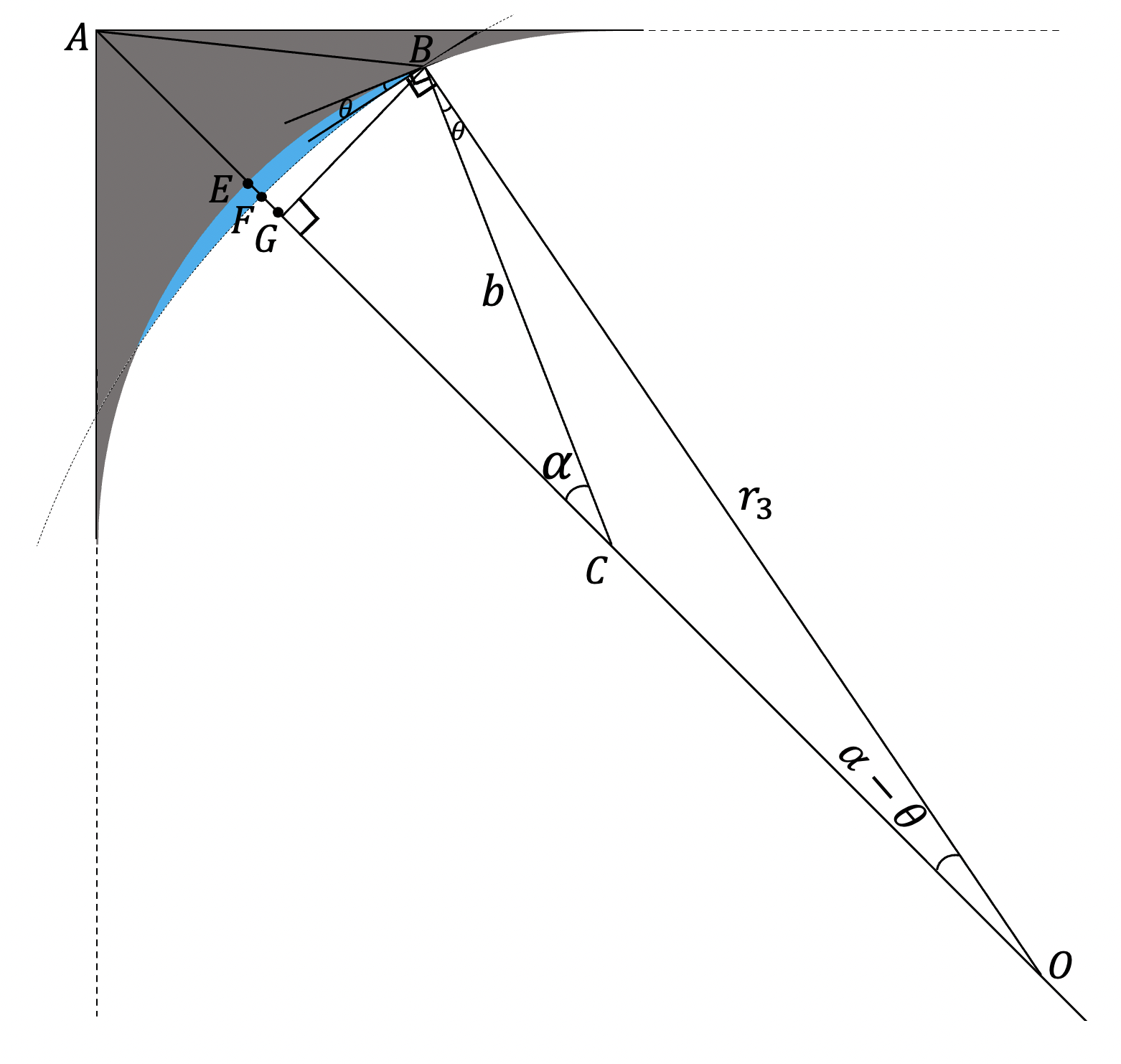}
  \caption{Square tube with rounded corners, case III with $\displaystyle \theta < \alpha$. The meniscus is convex towards to the corner.}
  \label{fig:rounded_caseIII1}
\end{figure}

For rounded corner, there is a case III when the saturation is small. 
The meniscus touches the solid surface on the rounded corner instead of the side-walls of the square (Fig. \ref{fig:rounded_caseIII1}). 
The meniscus $\overset{\frown}{\rm BF}$ is an arc with open angle $\alpha - \theta$ and radius $r_3$. 
Comparing $\triangle{\rm CBG}$ and $\triangle{\rm OBG}$
\begin{align}
	& \overline{\rm BG} = b \sin\alpha = r_3 \sin(\alpha-\theta) \\
	\Rightarrow \quad & r_3 = b \frac{\sin\alpha}{\sin(\alpha-\theta)}
\end{align}

From the geometry
\begin{align}
	S_{{\rm O}\overset{\frown}{\rm BF}} & = \frac{1}{2} (\alpha - \theta) r_3^2 \\
	S_{\triangle{\rm OBC}} & = \frac{1}{2} r_3 b \sin\theta \\
	S_{{\rm C}\overset{\frown}{\rm BF}} & =  S_{{\rm O}\overset{\frown}{\rm BF}} - S_{\triangle{\rm OBC}} = \frac{1}{2} (\alpha - \theta) r_3^2 - \frac{1}{2} r_3 b \sin\theta \\
	S_{{\rm C}\overset{\frown}{\rm BE}} & = \frac{1}{2} \alpha b^2 \\
	S_{\rm BEF} & = S_{{\rm C}\overset{\frown}{\rm BE}} - S_{{\rm C}\overset{\frown}{\rm BF}} \nonumber \\
	& = \frac{1}{2} b^2 \left[ \alpha - (\alpha-\theta) \frac{\sin^2\alpha}{\sin^2(\alpha-\theta)} + \frac{\sin\alpha \sin\theta}{\sin(\alpha-\theta)} \right]
\end{align}
We define a new function $C(\theta, \alpha)$ 
\begin{equation}
	\label{eq:C}
	C(\theta, \alpha) = \alpha - (\alpha-\theta) \frac{\sin^2\alpha}{\sin^2(\alpha-\theta)} + \frac{\sin\alpha \sin\theta}{\sin(\alpha-\theta)} \,.
\end{equation}
From $S_0 = 4a^2 B_2(\beta)$, we obtain the saturation as a function of the intermediate variable $\alpha$
\begin{equation}
	s = \frac{8 \times S_{\rm BEF}}{S_0} = \frac{ \beta^2 C(\alpha, \theta)}{B_2(\beta)} 
\end{equation}
The range of $\alpha$ is $\displaystyle 0 \le \alpha \le \frac{\pi}{4}$, which leads to an upper bound for $s$ in case III
\begin{equation}
	\label{eq:sc4}
	0 < s < s_{\rm c4}, \quad s_{\rm c4} = \frac{\beta^2 C(\frac{\pi}{4}, \theta)}{B_2(\beta)} \,.
\end{equation}
Comparing Eqs. (\ref{eq:sc3}) and (\ref{eq:sc4}), one can show that $s_{\rm c4} = s_{\rm c3}$. 
The energy curve $f(s)$ changes continuously from case III to case II. 

The length of solid-liquid interface is $\mathcal{L}_{\rm SL} = 8 \times \overset{\frown}{\rm BE} = 8 \alpha b$, and the length of liquid-vapor interface is $\mathcal{L}_{\rm LV} = 8 \times \overset{\frown}{\rm BF} = 8 (\alpha-\theta) r_3$. 
The interfacial energy is
\begin{align}
	f(s) & = \mathcal{L}_{\rm SL} (-\gamma\cos\theta) + \mathcal{L}_{\rm LV} \gamma \nonumber \\
	& = 8 \alpha b (-\gamma \cos\theta) + 8 (\alpha-\theta) r_3  \gamma \\
	\frac{f(s)}{a\gamma} &= 8 \beta \left[ - \alpha \cos\theta + (\alpha - \theta) \frac{ \sin\alpha }{\sin(\alpha-\theta)} \right]
	\label{eq:fs_rIII1}
\end{align}

\begin{figure}[htbp]
  \includegraphics[width=0.5\columnwidth]{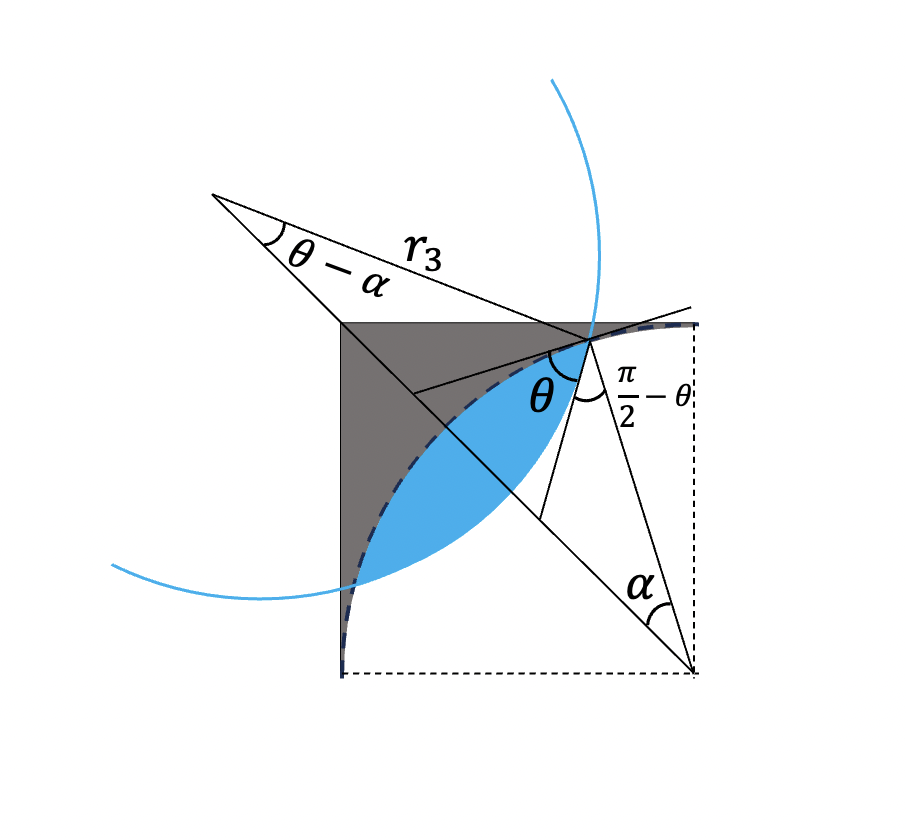}
  \caption{Square tube with rounded corners, case III with $\displaystyle \theta > \alpha$. The meniscus is convex towards to the tube center.}
  \label{fig:rounded_caseIII2}
\end{figure}

The above derivation only applies when $\theta < \alpha$. 
For the opposite case of $\theta > \alpha$ (Fig. \ref{fig:rounded_caseIII2}), the saturation and the interfacial energy have the same form. 

\subsection{Summary of rounded corner $f(s)$}

We summarize the interfacial energy functions for rounded corners.
\begin{itemize}

\item {\bf case I}: $\displaystyle s_{\rm c1} \le s \le 1$,  
$s_{\rm c1} = 1 - \frac{\pi}{4 B_2(\beta)}$
\begin{equation}
	\frac{f(s)}{a\gamma} = - 8 \cos\theta B_1(\beta) + 4 \sqrt{\pi B_2(\beta)} (1-s)^{1/2}
\end{equation}

\item {\bf case II}: $s_{\rm c3} \le s \le s_{\rm c2}$,
\begin{align}
  s_{\rm c2} &= \frac{ A(\theta) (\cos\theta - \sin\theta)^{-2} - (1-B_2(\beta)) }{ B_2(\beta) } \nonumber \\
  s_{\rm c3} &= \frac{ \beta^2 A(\theta) (\cos\theta - \sin\theta)^{-2} - (1-B_2(\beta)) }{ B_2(\beta) } \nonumber \\
  \frac{f(s)}{a\gamma} &= 8(1-B_1(\beta)) \cos\theta - {\rm sign} \left( \frac{\pi}{4}-\theta \right) 8 \sqrt{A(\theta)} \sqrt{ B_2(\beta) s + (1-B_2(\beta)) }
\end{align}

\item {\bf case III}: $0 \le s \le s_{\rm c3}$ and $\displaystyle 0 < \alpha < \frac{\pi}{4}$
\begin{align}
	s(\alpha) &= \frac{ \beta^2 C(\alpha, \theta)}{B_2(\beta)} \\
	\frac{f(s(\alpha))}{a\gamma} &= 8 \beta \left[ - \alpha \cos\theta + (\alpha - \theta) \frac{ \sin\alpha }{\sin(\alpha-\theta)} \right]
\end{align}

\end{itemize}
$A(\theta)$, $B_1(\beta)$, $B_2(\beta)$, and $C(\alpha, \theta)$ are given in Eqs. (\ref{eq:A}), (\ref{eq:B1}), (\ref{eq:B2}), and (\ref{eq:C}), respectively.

\section{Construction of phase diagrams}
\label{app:pd}

To determine $s^*$ and $s^{**}$, we only need the energy function of case II (\ref{eq:fs_rII1})
\begin{equation}
	\frac{f(s)}{a\gamma} = 8(1-B_1) \cos\theta - 8 \sqrt{A} \sqrt{ B_2 s + (1-B_2) } \,.
\end{equation}
To simplify the notation, we have omit the dependence of $A(\theta)$ and $B_{1,2}(\beta)$. 
For a given pair of $(\theta, \beta)$, they are constants. 

The gradient is 
\begin{equation}
	\frac{f'(s)}{a\gamma} = - 4 \sqrt{A} B_2 \left( B_2 s + (1-B_2 ) \right)^{-1/2} \,.
\end{equation}

\subsection{calculation of $s^*$}

The condition for $s^*$ is 
\begin{equation}
	f'(s) = \frac{f(1)-f(s)}{1-s} \,,
\end{equation}
which leads to 
\begin{equation}
	\frac{ - 4 A B_2 }{ \sqrt{A} \sqrt{B_2 s + (1-B_2)} } = \frac{ - 8 \cos\theta + 8 \sqrt{A} \sqrt{B_2 s + (1-B_2)} }{1-s}
\end{equation}
Let $x=\sqrt{A} \sqrt{B_2 s + (1-B_2)}$, $x$ satisfies the equation
\begin{align}
	& x^2 - 2\cos\theta x + A = 0  \\
	\Rightarrow \quad & x = \cos\theta - \sqrt{\cos^2\theta - A} \,.
\end{align}
We chose the solution with negative sign. 
Once $x$ is obtained, we can use the reverse relation to get $s^*$
\begin{equation}
	s^* = \frac{ x^2 - A ( 1- B_2)}{A B_2}
\end{equation}

\begin{figure}[htp]
  \includegraphics[width=0.9\columnwidth]{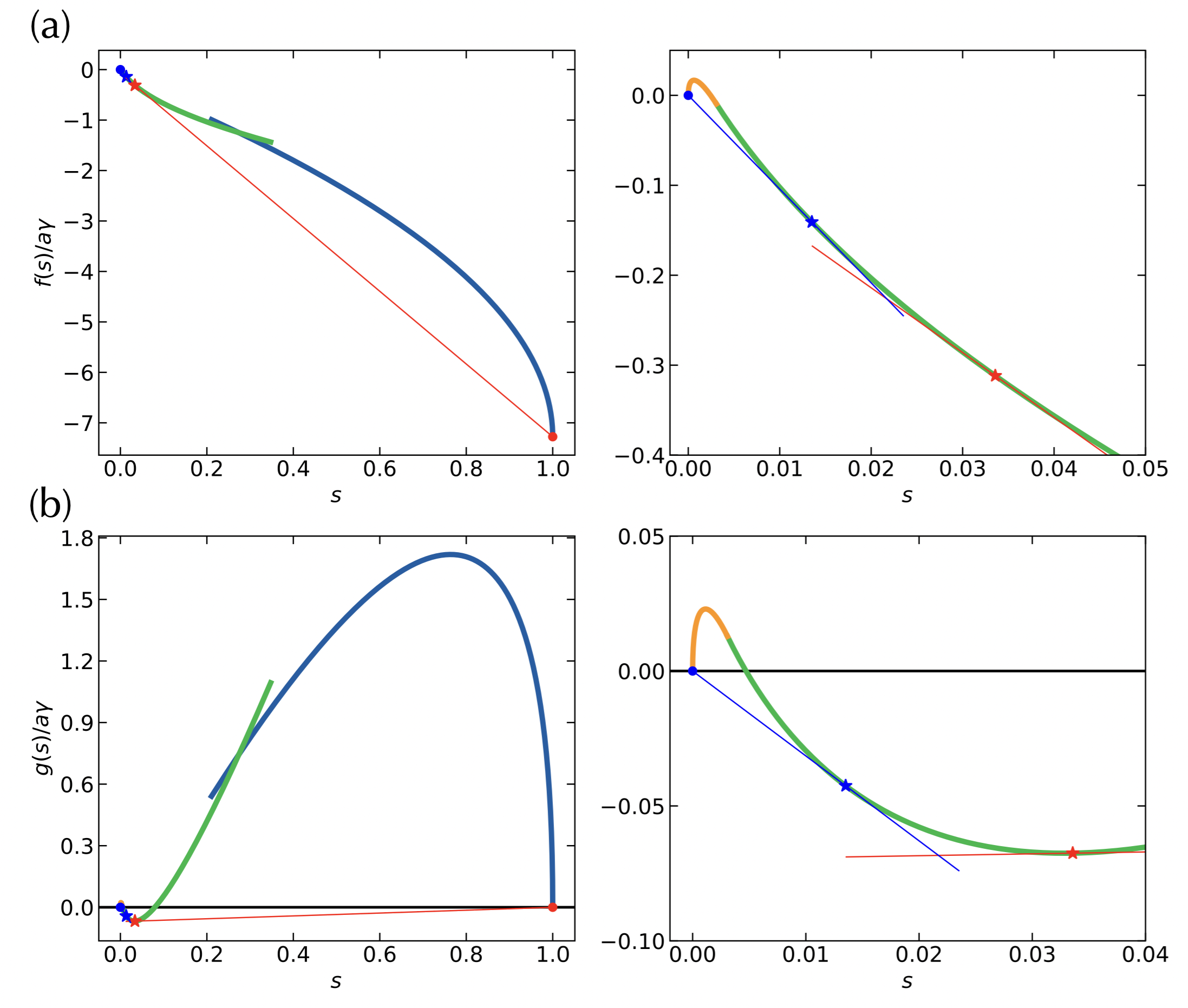}
  \caption{Square tube with rounded corners, interfacial energy curves $f(s)$ and $g(s)$. Contact angle $\theta=20^{\circ}$ and $\beta=0.1$. }
  \label{fig:freeE_rounded1}
\end{figure}

\begin{figure}[htp]
  \includegraphics[width=0.9\columnwidth]{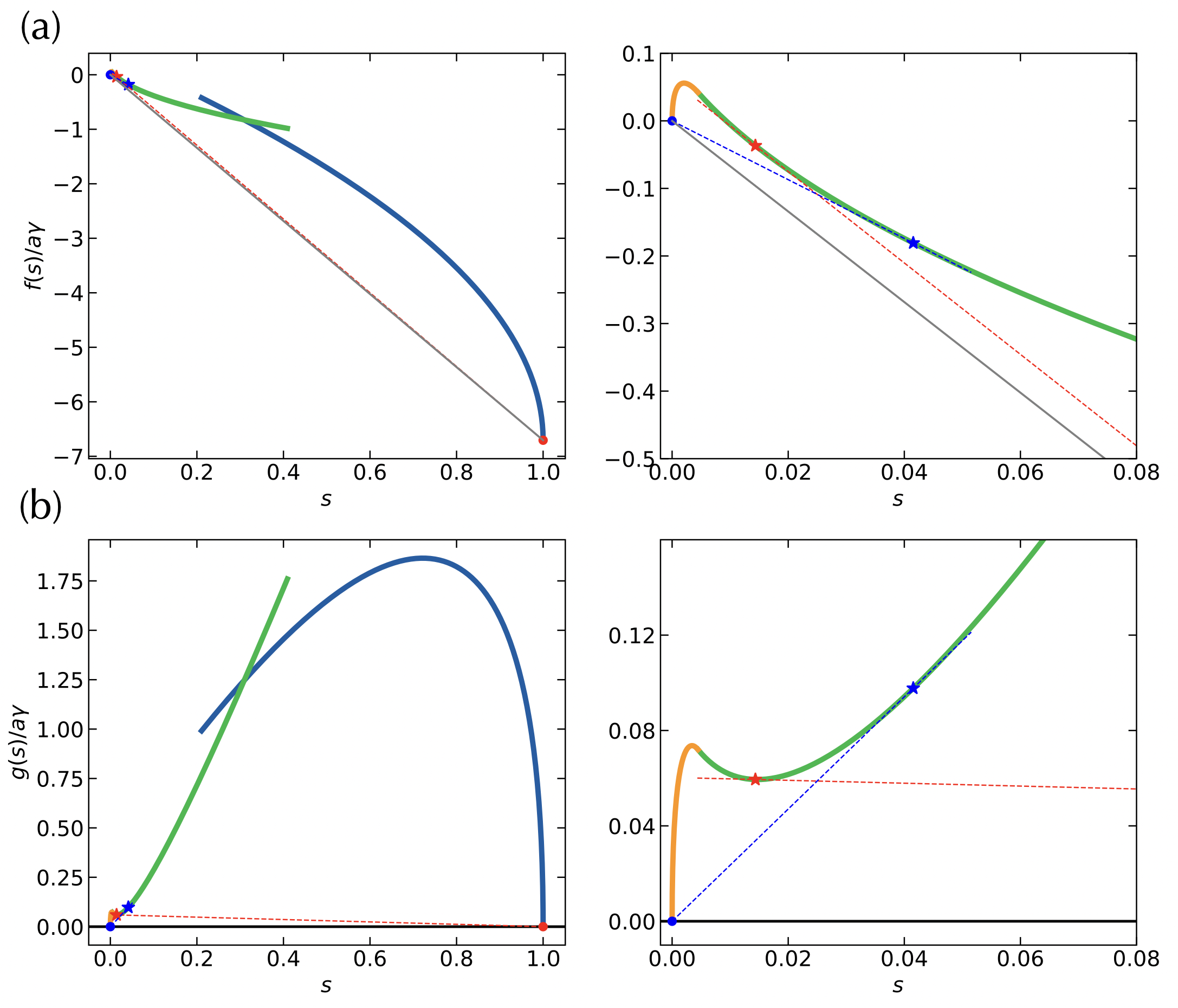}
  \caption{Square tube with rounded corners, interfacial energy curves $f(s)$ and $g(s)$. Contact angle $\theta=30^{\circ}$ and $\beta=0.15$.}
  \label{fig:freeE_rounded2}
\end{figure}

\subsection{calculation of $s^{**}$}

The condition for $s^{**}$ is 
\begin{equation}
	f'(s) = \frac{f(s)}{s} \,,
\end{equation}
which leads to 
\begin{equation}
	\frac{ - 4 A B_2 }{ \sqrt{A} \sqrt{B_2 s + (1-B_2)} } = \frac{ 8(1-B_1) \cos\theta + 8 \sqrt{A} \sqrt{B_2 s + (1-B_2)} }{s}
\end{equation}
Let $y=\sqrt{A} \sqrt{B_2 s + (1-B_2)}$, $y$ satisfies the equation
\begin{align}
	& y^2 - 2(1-B_1)\cos\theta y + A(1-B_2) = 0  \\
	\Rightarrow \quad & y = (1-B_1) \cos\theta - \sqrt{(1-B_1)^2 \cos^2\theta - A(1-B_2)^2} \,.
\end{align}
We again chose the solution with negative sign. 
Once $y$ is obtained, we can use the reverse relation to get $s^{**}$
\begin{equation}
	s^{**} = \frac{ y^2 - A ( 1- B_2)}{A B_2}
\end{equation}

Figure \ref{fig:freeE_rounded1} shows the interfacial energy $f(s)$ and modified one $g(s)$ for $\theta=20^{\circ}$ and $\beta=0.15$. 
Both $s^*$ and $s^{**}$ are present and $s^* > s^{**}$. 

Figure \ref{fig:freeE_rounded2} shows the interfacial energy $f(s)$ and modified one $g(s)$ for $\theta=30^{\circ}$ and $\beta=0.15$. 
In this case, $s^{**} > s^{*}$, therefore fingers are absent.

\begin{figure}[htp]
  \includegraphics[width=1.0\columnwidth]{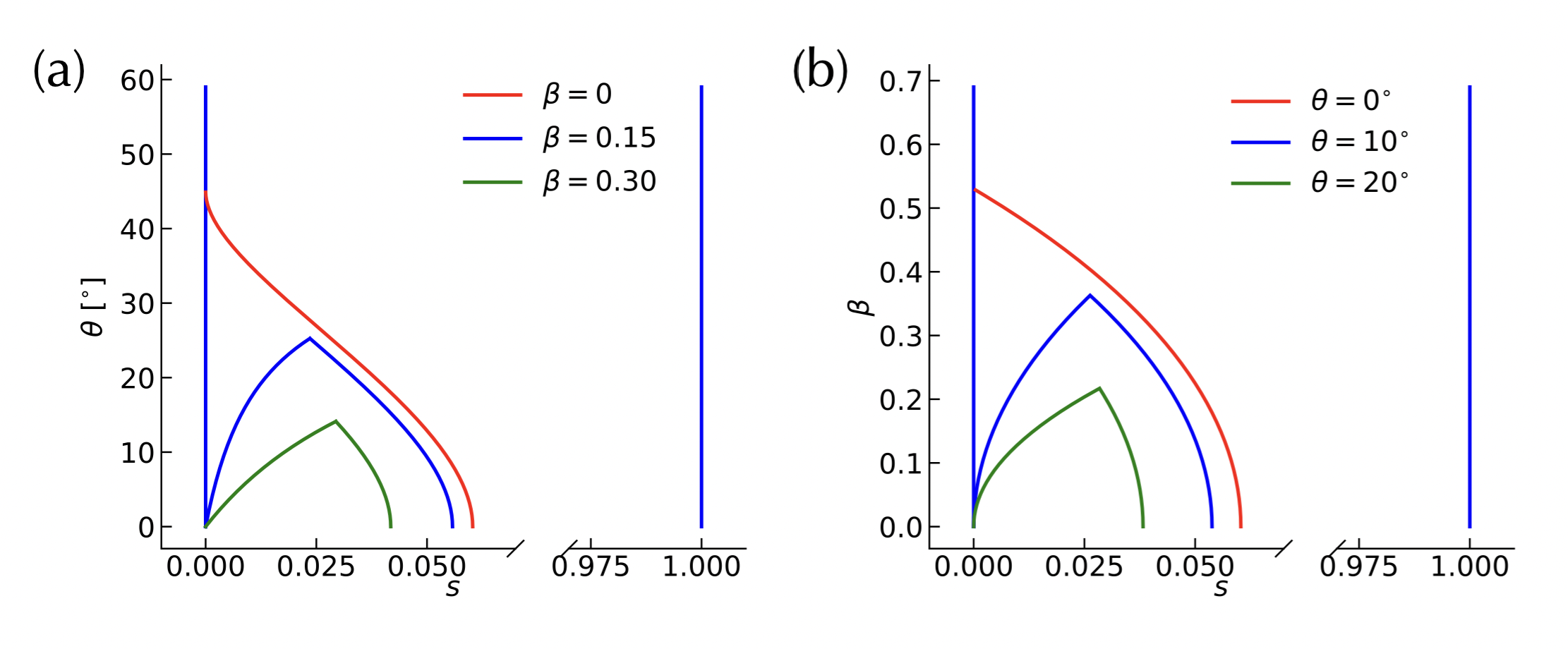}
  \caption{(a) Phase diagram for a square tube with rounded corners in the plane of average saturation and contact angle (${s}$-$\theta$). (b) Phase diagram in the plane of average saturation and roundness (${s}$-$\beta$). }
  \label{fig:pd_sat}
\end{figure}

Once the value of $s^*$ and $s^{**}$ are obtained, we can construct the phase diagram. 
Figure \ref{fig:pd_sat}(a) shows the phase diagram in the plane of saturation and contact angle ($s$-$\theta$) for different roundness. 
The left curve is $s^{**}$ and the right curve is $s^{*}$. 
The $s^{**}$ curve vanishes when $\theta = 0$ and the $s^{*}$ touches the $s=0$ axis at $\theta=45^{\circ}$. 
This indicates the disappearance of finger-dry coexistence. 

Figure \ref{fig:pd_sat}(b) shows a similar phase diagram, but constructed in the plane of saturation and roundness ($s$-$\beta$) for different contact angles. 
The $s^{**}$ curve vanishes when $\beta = 0$ and the $s^{*}$ touches the $s=0$ axis at $\beta \simeq 0.53$. 

\subsection{$s^*=s^{**}$ condition}

The condition for $s^* = s^{**}$ is more clearly seen in the modified energy $g(s)$. 
As shown in Fig. \ref{fig:freeE_rounded1}(c), $s^{**}<s^{*}$ corresponds to the minimum of $g(s)$ is less than zero. 
When $s^{**} > s^{*}$, we have $g(s) \big|_{\rm min} > 0$. 
Then the boundary of finger is determined by $g(s) \big|_{\rm min} = 0$. 

The modified energy is
\begin{align}
	g(s) & = f(s) - f(1) s \nonumber \\
	\frac{g(s)}{a\gamma} & = 8(1-B_1) \cos\theta - 8 \sqrt{A} \sqrt{B_2 s + (1-B_2)} + 8 \cos\theta B_1 s \,.
\end{align}
The gradient is 
\begin{equation}
	\frac{g'(s)}{a\gamma} = - \frac{4 \sqrt{A} B_2}{\sqrt{B_2 s + (1-B_2)}} + 8 \cos\theta B_1 \,.
\end{equation}

The location of minimum $s_{\rm min}$ is determined by $g'(s)=0$, which leads to 
\begin{equation}
	s_{\rm min} =  \frac{AB_2}{(2B_1 \cos\theta)^2 } -  \frac{1-B_2}{B_2}  \,.
\end{equation}
Substituting above expression into $g(s)$
\begin{equation}
	\frac{g(s_{\rm min})}{a\gamma} = 8(1-B_1)\cos\theta - \frac{2AB_2}{B_1\cos\theta} - \frac{8 \cos\theta B_1(1-B_2)}{B_2} \,.
\end{equation}
Setting this value to be zero, we obtain the following equation
\begin{equation} 
	A \left( \frac{B_2}{B_1} \right)^2 - 4 \cos^2\theta \left( \frac{B_2}{B_1} \right) + 4 \cos^2 \theta = 0 \,.
\end{equation}
Let $z=B_2/B_1$, we can obtain the solution of the above equation
\begin{equation}
	z = \frac{ 2 \cos^2 \theta - \sqrt{4 \cos^4 \theta - 4 A \cos^2\theta} }{A} \,.
	\label{eq:z1}
\end{equation}

The critical roundness $\beta_{\rm crit}$ satisfies 
\begin{align}
	& z = \frac{B_2}{B_1} = \frac{ 1 - \beta^2 \left( 1 - \frac{\pi}{4} \right)}{ 1 - \beta \left( 1 - \frac{\pi}{4} \right)} \\
	\Rightarrow \quad & \beta^2 - z \beta + \frac{z-1}{1- \frac{\pi}{4}} = 0
\end{align}
The solution is
\begin{equation}
	\beta_{\rm crit} = \frac{ z - \sqrt{z^2 - \frac{4(z-1)}{1-\frac{\pi}{4}}} }{2} \,.
	\label{eq:z2}
\end{equation}
For fixed value of contact angle $\theta$, we can calculate $z$ from Eq. (\ref{eq:z1}). 
Substituting $z$ into Eq. (\ref{eq:z2}) leads to $\beta_{\rm crit}$. 
The phase diagram Fig. \ref{fig:pd_beta_theta} then can be constructed.

For the special case $\theta=0$, $\displaystyle A(0) = 1 - \frac{\pi}{4}$, the equation for $z$ becomes
\begin{equation} 
	(1 - \frac{\pi}{4}) z^2 - 4 z + 4 =0 \,.
\end{equation}
The solution is $z=(8-4\sqrt{\pi})/(4-\pi)$. 
This leads to $\beta_{\rm crit} = (4-2\sqrt{\pi})/(4-\pi) \simeq 0.53$.

\end{document}